\documentclass{article}
\usepackage{arxiv}

\usepackage[utf8]{inputenc} % allow utf-8 input
\usepackage[T1]{fontenc}    % use 8-bit T1 fonts
\usepackage{hyperref}       % hyperlinks
\usepackage{url}            % simple URL typesetting
\usepackage{booktabs}       % professional-quality tables
\usepackage{amsfonts}       % blackboard math symbols
\usepackage{microtype}      % microtypography
\usepackage{graphicx}
\usepackage{natbib}

\usepackage{amssymb}
\usepackage{amsmath}
\usepackage{mathtools}
\usepackage{placeins}
\usepackage{natbib}
\usepackage{caption}
\usepackage{verbatim}
\usepackage{rotating}
\usepackage[utf8]{inputenc}
\usepackage{pgfplots}
\usepgfplotslibrary{groupplots}

\pgfplotsset{compat=newest} 
\newcommand{\vx}{\boldsymbol{x}}
\newcommand{\vu}{\boldsymbol{u}}

\newcommand{\vX}{\boldsymbol{X}}
\newcommand{\vy}{\boldsymbol{y}}

\newcommand{\dd}{\boldsymbol{d_d}}
\newcommand{\cm}{\mathcal{M}}

\newcommand{\ch}{\mathcal{H}}

\newcommand{\ve}[1]{\boldsymbol{#1}}

\title{Bayesian tomography with prior-knowledge-based parametrization and surrogate modeling}

\author{ Giovanni Angelo ~Meles \\
	Institute of Earth Sciences \\
	University of Lausanne\\
	CH-1015 Lausanne  \\
	\texttt{Giovanni.Meles@unil.ch} \\
	\And
	Niklas ~Linde \\
	Institute of Earth Sciences \\
	University of Lausanne\\
	CH-1015 Lausanne  \\
	\texttt{Niklas.Linde@unil.ch} \\
		\And
	Stefano ~Marelli \\
	Department of Civil, Environmental and Geomatic Engineering \\
	ETH Zurich\\
	CH-8093 Zürich \\
	\texttt{Marelli@ibk.baug.ethz.ch} \\
}

\renewcommand{\shorttitle}{Bayesian tomography with prior-knowledge-based parametrization and surrogate modeling}

\hypersetup{
pdftitle={A template for the arxiv style},
pdfsubject={q-bio.NC, q-bio.QM},
pdfauthor={David S.~Hippocampus, Elias D.~Striatum},
pdfkeywords={First keyword, Second keyword, More},
}

\begin{document}
\maketitle

\begin{abstract}
We present a Bayesian tomography framework operating with prior-knowledge-based parametrization that is accelerated by surrogate models. Standard high-fidelity forward solvers (e.g.,  Finite-Difference Time-Domain schemes) solve wave equations with natural spatial parametrizations based on fine discretization. Similar parametrizations, typically involving tens of thousand of variables, are usually employed to parameterize the subsurface in tomography applications.
When the data do not allow to resolve details at such finely parameterized scales, it is often beneficial to instead rely on a prior-knowledge-based parametrization defined on a lower dimension domain (or manifold). Due to the increased identifiability in the reduced domain, the concomitant inversion is better constrained and generally faster.
% CHECK WITH NIKLAS AND STEFANO:
%Since the cost of practical implementation of Markov-Chain Monte Carlo (MCMC) schemes can be significantly affected by the parameter dimension, a dimensionality reduction could also be beneficial in terms of computation demands.
We illustrate the potential of a prior-knowledge-based approach by considering ground penetrating radar (GPR) travel-time tomography in a crosshole configuration. An effective parametrization of the input (i.e., the permittivity distributions determining the slowness field) and compression in the output (i.e., the travel-time gathers) spaces are achieved via data-driven principal component decomposition based on random realizations of the prior Gaussian-process model with a truncation determined by the performances of the standard solver on the full and reduced model domains. 
To accelerate the inversion process, we employ a high-fidelity polynomial chaos expansion (PCE) surrogate model. We investigate the impact of the size of the training set on the performance of the PCE and show that a few hundreds design 
data sets is sufficient to provide reliable Markov chain Monte Carlo inversion at a fraction of the cost associated with a standard approach involving a fine discretization and physics-based forward solvers. Appropriate uncertainty quantification is achieved by reintroducing the truncated higher-order principle components in the original model space after inversion on the manifold and by adapting a likelihood function that accounts for the fact that the truncated higher-order components are not completely located in the null-space.
\end{abstract}

% keywords can be removed
\keywords{ Bayesian inversion, surrogate modelling, tomography, GPR.
}

\section{Introduction}

Travel-time tomography refers to a large class of noninvasive imaging methods with applications comprising natural resource exploration \citep{taillandier2009first}, extraterrestrial investigations \citep{zhao2008seismic,khan2016single}, medical diagnosis \citep{stotzka2002medical} and non-destructive testing \citep{tant2018transdimensional}. In travel-time tomography, active or passive sources generate acoustic, elastic or electromagnetic waves, and the corresponding  first-arrival times are used to infer properties of the illuminated medium \citep{rawlinson2006simultaneous}. 
High-resolution travel-time tomography of the shallow subsurface can be performed using ground-penetrating radar (GPR).
GPR wave-propagation depends on the distribution of dielectric permittivity ($\epsilon$) and electric conductivity ($\sigma$) in the subsurface. Permittivity and conductivity fields mainly determine wave propagation velocity and attenuation, respectively, while variations of either property result in scattering \citep{balanis2012advanced}.
The penetration depth range varies greatly, ranging from few centimeters in wet bentonite clay up to several kilometers in ice, while in dry environments GPR waves can typically penetrate up to 30 meters \citep{slob2010surface}. 
Under ideal conditions in terms of coverage, signal-to-noise ratio and processing, GPR data can be used to obtain high-resolution estimates of permittivity and conductivity distributions, while simpler configurations and analyses allow for inexpensive identification/localization of subsurface discontinuities/anomalies \citep{grasmueck2005full,dou2016real}. 
Reflection surveys are relatively easy and inexpensive to deploy, but their performance is heavily affected by the limited penetration depth of GPR waves in conductive media \citep{piscitelli2007gpr}, such as soils. Cross-hole configurations are more demanding in terms of implementation as boreholes are needed, but are well suited for groundwater investigations \citep{labrecque2002three,annan2005gpr}.

Deterministic GPR tomography can be carried out either by considering the travel-time and/or the amplitude \citep{olsson1992borehole} of the first-arrival signals, or even considering the waveforms of the measured electromagnetic (EM) fields \citep{ernst2007full,Kuroda2007}, so-called full waveform inversion (FWI). %Notwithstanding the  sophistication of deterministic inversion algorithms, GPR tomography tends to be ill-posed \citep{mangue2009first}, and regularization methods are often needed to stabilize the inversion \citep{tikhonov1977solutions,aster2018parameter}.
Probabilistic inversion approaches based on Bayesian statistics have been proposed as an alternative, both for travel-time and FWI inversion \citep{nielsen2010estimation,cordua2012monte,hunziker2019bayesian}. A drawback of such search methods is the high computational cost associated with the many (often hundreds of thousands) solutions of the forward problem required to locate and sample from the posterior distribution; this can be computationally prohibitive even when state-of-the-art solvers are used \citep{warren2019cuda}. As an alternative to physics-based solvers, machine learning (ML) tools have become increasingly popular in recent years. Machine learning offers a number of approaches to obtain inexpensive proxies mimicking the input-output relations of standard forward solvers, and regressive neural networks have already been successfully used in the implementation of various  global search GPR algorithms \citep{hansen2017efficient,giannakis2019machine}.
In most inversion schemes, the input domain is parametrized relying on the same spatial discretization as the one used in standard forward solvers. 
When Finite-Difference Time-Domain (FDTD) schemes are employed, regular grids are usually considered and each grid point represents one inversion parameter. 
Standard 2D problems can then easily involve tens of thousands of parameters, leading to computationally extreme demands when relying on a Bayesian formulation of the inverse problem. Various strategies to reduce the effective dimensionality of the model domain have been attempted.
For example, since the  resolving power (in terms of wavelength) associated with the data is  much  lower  than  the  discretization  needed  for  accurate forward  modelling, adjacent cells used for forward modeling can be clustered to represent a single inversion parameter. 
Such compression reduces  the  memory  requirements  of deterministic inversion schemes \citep{ernst2007full} and decreases the ill-posedness of the corresponding problem, but parameterizations should ideally rather be based on available prior knowledge about the unknowns rather than modelling-specific constraints \citep{haario2004markov}.
%Here, we propose to use a different type of parametrization, guided by available information and use it to perform Bayesian inversion.
Here we pursue an approach in which the effective dimensionality of the problem is determined by comparison of predictions of the the standard forward solver in the full model domain with respect to prior-knowledge-based parameters defining a lower-dimensional model domain (a manifold), such that the associated modeling errors are well below the expected noise level of the data. 

To further decrease the computational costs associated with Markov chain Monte Carlo (MCMC) inversion, we employ unbiased surrogate modelling to approximate standard travel-time forward solvers  \citep{xiu2002wiener}.
Various classes of surrogate models, such as those based on Gaussian process models or kriging
\citep{sacks1989designs,rasmussen2003gaussian} and polynomial chaos expansions (PCEs) \citep{xiu2002wiener,blatman2011adaptive}, can be employed in Bayesian inverse problems \citep{nagel2019bayesian,higdon2015bayesian,marzouk2009stochastic,marzouk2007stochastic,wagner2020bayesian,wagner2021bayesian}. In this contribution, we rely on regression-based sparse PCE due to their extrapolation capabilities and robustness with respect to noise \citep{blatman2011adaptive,luethen2021sparse,marelli2021stochastic}. Several applications of PCE in the geosciences have been proposed. Examples include hydrogeological characterization of a three-dimensional multilayered aquifer, in which an efficient posterior exploration of a high-dimensional space is presented by \citet{laloy2013efficient},  where the Karhunen-Lo\`eve transform is used to significantly reduce the dimensionality of the parameter space. A PCE model is then introduced  to perform a two-stage MCMC inversion. PCE has also been used to automate velocity analysis in seismic processing, where the ideal velocity model can be defined as an optimizer of a variational integral in a semblance field. Calculating the relevant variational integrals is expensive and using probabilistic methods  become computationally impractical. However, the variational integral can be replaced by a PCE, which results in massive reduction in overall computational costs \citep{abbasi2017polynomial}. PCE can also be used as a compression tool. In a magnetometric resistivity tomography study, the logarithm of the electrical conductivity field is compressed using few tens of Hermite polynomials rather than tens of thousands of discretization parameters, resulting in important computational gains \citep{vu2020magnetometric}.
%ADD: \citep{manassero2020reduced,ortega2020fast}

In the context of Bayesian travel-time tomography, a pioneering use of PCE surrogates for 1.5D media has been recently introduced by \citet{sochala2021polynomial}. In their work, \citet{sochala2021polynomial} demonstrate the potential of PCE trained using eikonal solutions in Bayesian travel-time tomography, assuming that the observation error dominates other sources of uncertainty. 
Here, we propose a Bayesian framework based on PCE to perform travel-time tomography of more complex 2D media. We show that PCE trained with FDTD data can be successfully designed and efficiently used for Bayesian inversion. Crucially, we also show that superior convergence properties are achieved when modelling error is taken into account in the inversion process.

\section{Methodology}

\subsection{Bayesian inversion}
\label{sBayes}
Forward models are mathematical tools used to predict the outcome of physical experiments by operating on a set of input parameters. A forward problem expresses the relationship between input parameters and output values:
\begin{equation}
 \mathcal{F}(\vu) = \vy + \epsilon,
 \label{forward}
\end{equation}
with $\mathcal{F}$, $\vu$, $\vy$ and $\epsilon$ standing for the physical law or forward operator, typically involving PDEs acting on locally defined quantities, the input parameters, the output and a noise term, respectively.
The corresponding inverse problem aims at inferring properties of $\vu$ given data $\vy$ and any available prior information about $\vu$.
A general solution to an inverse problem is the posterior distribution defined over the input domain of the corresponding forward problem. This solution can be formally expressed using Bayes' theorem:
\begin{equation}
 P(\vu|\vy)= \frac{P(\vy|\vu)P(\vu)}{P(\vy)}
 \label{Bayes}=\frac{L(\vu)P(\vu)}{P(\vy)},
\end{equation}where $P(\vu|\vy)$ is the posterior distribution of the input parameter $\vu$ given the data $\vy$, $P(\vy|\vu)$ (also indicated as $\mathcal{L}(\vu)$) expresses the probability of observing the data $\vy$ given the input parameter $\vu$, $P(\vu)$ the prior distribution in the input parameter domain and $P(\vy)$ the marginalized likelihood with respect to the input parameters (also known as evidence). While formally exact, Eq.~\eqref{Bayes} is seldom used when solving inverse problems as the computation of the evidence is in most cases very expensive. Instead,  since the product $\mathcal{L}(\vu)P(\vu) $ is proportional to the posterior distribution and can be calculated for any value $\vu$, one can use a Markov chain Monte Carlo (MCMC) method to draw samples from  $P(\vu|\vy)$ without considering the evidence. 
%Evaluation of the prior distribution $P(\vu)$ can be \textit{theoretically} achieved for example via random realizations of a generative model embedding the available prior knowledge. However, actual computation of $P(\vu)$ can be problematic when $\vu$ is high-dimensional. %Note also that physics-based coordinates are local, whereas prior information is typically provided at a non-local scale.
%Moreover, $\mathcal{L}(\vu)P(\vu) $ needs to be determined up to hundreds of thousands of times to achieve steady-state convergence in MCMC \citep{gelman1992inference} which can be unfeasible for expensive models $\mathcal{F}(\vu) $.
%For tomography problems $\vu$ is typically representing the slowness distribution at every point/pixel in space, and therefore proper evaluation of $P(\vu)$ can be unfeasible. 
Here, we extract the most relevant features from the original input domain and perform the inversion on an effective lower-dimensional manifold (see subsection \ref{effective}). We also reduce the often excessive computational burden of high-dimensional MCMC problems by approximating the forward model via a much less demanding PCE surrogate  (see subsection \ref{PCEtheory}) while accounting for this approximation in the inversion. 

\subsection{Identification of an effective manifold}
\label{effective}
Without any loss of generality, the input in Eq. ~\eqref{forward}, can be formally expressed using new variables via a standard change of coordinates:
\begin{equation}
 \mathcal{F}(\vu({\vx_{full}})) = {\vy} +\epsilon.
 \label{sforward}
\end{equation}
Such a reparameterization can help to shift the focus of the model representation from the requirements of the forward solver (local fine discretization) to one that is better suited for the Bayesian inverse framework in the sense that it encodes the prior distribution. The new coordinates $\vx_{full}$ can be defined, for example, in terms of Fourier coefficients, autoencoders, or principal components associated with random realizations of the generative model, which in turn can be related to the stochastic/topological properties inherent to the prior knowledge. 
Perhaps the most natural and widely used example is to discretize geostatistical prior knowledge on geostatistical hyper parameters into the Karhunen-Loève modes of a corresponding Gaussian random field.
In this contribution, we rely instead on a data-driven approach using samples from a pre-existing generative model as this makes the method insensitive to any assumptions of the prior being a multi-variate Gaussian. In particular, we extract principal components (PCA) from a set of realizations of the prior model to identify the relevant non-local features of the input. 
Other transforms are possible provided it can be projected (transformed) on the forward-solver mesh.

In the case of PCA considered here, the forward problem can be exactly expressed as operating directly on the new set of coordinates, that is, the principal component coefficients:
\begin{equation}
 \mathcal{M}({\vx_{full}}) = {\vy} +\epsilon,
 \label{forwardcoordinates}
\end{equation}
where $\mathcal{M}=\mathcal{F\circ\vu}$, and $\circ$ stands for function composition. 
Under the additional key assumption that the model can be faithfully represented by operating on an effective truncated $M$-dimensional subset $\vx$ of the new coordinates $\vx_{full}$, we can write:
\begin{equation}
 \mathcal{M}({\vx}) = {\vy} +\hat{\epsilon},
 \label{forwardreduced}
\end{equation}
where $\hat{\epsilon}$ is a noise term including both observational noise and modeling errors related to the projection to the manifold. 
We assume that the first $M$ components of ${\vx}$ include the most relevant aspects of $\vx_{full}$, implying that we ignore the remaining coefficients by setting them to zero. However, we will include this error in the inversion by using an appropriate likelihood function that considers this error and other modeling errors induced by the surrogate modeling. 
Introducing a model operating on an effective set of coordinate ${\vx}$ casts the inverse problem on the $M$-dimensional manifold as:
\begin{equation}
 P(\vx|\vy)= \frac{P(\vy|\vx)P(\vx)}{P(\vy)}.
 \label{BayesReduced}
\end{equation}
When $\vx$ is low-dimensional, the corresponding statistical prior distribution $P(\vx)$  can be \textit{practically} estimated using random realizations of the generative model of $P(\vu)$. 
 Many different MCMC sampling techniques could be used to draw samples from the posterior distribution $P(\vx|\vy)$. Here, we use a Metropolis-Hastings algorithm based on a Gaussian proposal distribution with proposal scales
determined by a diagonal covariance matrix. Using one common scaling factor, each element along the diagonal is proportional to the prior marginal variance associated with the corresponding input parameter  \citep{hastings1970monte,marelli2014uqlab}.
Since we assume that the impact of the ignored higher-order variables is non-zero, elements from this 'null space' defined by their prior distributions can be added after the inversion and used to properly asses uncertainty on $\vu$.

A compressing procedure, consisting of a change of coordinates and a truncation, can also be made a-priori to reduce the dimensionality of the output domain. Note that the benefit of using compression in the output domain is associated with gains in storage and computational efficiency since each output variable requires its own PCE expansion (see section \ref{parametrization}).
\subsection{Polynomial chaos expansion }
\label{PCEtheory}
Forward models $\mathcal{M}$ are typically implemented by potentially expensive schemes (e.g. finite element methods, FDTD, etc.) in order to accurately mimic the relationship between $\vx$ and $\vy$ in Eq.~\eqref{forwardreduced}. 
A surrogate model $\tilde{\mathcal{M}}$  is a function that seeks to emulate the behaviour of an expensive forward model at negligible cost per run: 
\begin{equation}
\tilde{\mathcal{M}}(\vx) \approx \mathcal{M}(\vx).
 \label{surrogate}
\end{equation}

Among the many available classes of surrogate models, sparse adaptive polynomial chaos expansions are one of the most widely used due to their efficiency (the evaluation of the PCE polynomials can be done using sparse matrix multiplication), flexibility (e.g., in terms of polynomial definition and properties) and ease of deployment \citep{xiu2002wiener,blatman2011adaptive,luethen2021sparse,metivier2020efficient}.

Polynomial chaos expansions are a type of stochastic spectral expansions that can be used to project the forward operator onto an orthonormal polynomial basis of a suitable functional space.
Consider a stochastic Hilbert space $\ch$ equipped with the following scalar product:
\begin{equation}
    \label{eqn:scalar product}
    \langle g(\vx), h(\vx) \rangle \equiv \mathbb{E}_{\vX}\left[ g\cdot h\right] = 
    \int\limits_{\Omega_{\vX}}g(\vx)h(\vx) f_{\vX}(\vx)d\vx,
\end{equation}
where the $\mathbb{E}_{\vX}$ symbol denotes the statistical expectation operator with respect to the joint probabilistic distribution $f_{\vX}(\vx)$ of the random vector of the input parameters $\vX$.
Then any forward operator of finite variance $\mathcal{M}(\vx)\in \ch$, that is, 
\begin{equation}\label{eqn:squareInt}
\int\limits_{\Omega_{\vX}}\mathcal{M}(\vx)^2 f_{\vX}(\vx)d\vx \equiv \langle \cm(\vx), \cm(\vx) \rangle < +\infty ,
\end{equation}
can be expanded as a sum of polynomial basis elements $\Psi_{\boldsymbol{\alpha}}$ of $\ch$ as \citep{xiu2002wiener}:
\begin{equation}
 \mathcal{M}(\vx) = \sum_{\boldsymbol{\alpha} \in \mathbb{N}^{{M}}} a_{\boldsymbol{\alpha}} \Psi_{\boldsymbol{\alpha}}(\vx),
 \label{PCE_Total}
\end{equation}
where the multivariate polynomial basis functions $\Psi_{\boldsymbol{\alpha}}$ are orthonormal with respect to the inner product in Eq.~\eqref{eqn:scalar product}, that is:
\begin{equation}
    \label{eqn:ortho basis}
    \langle \Psi_{\boldsymbol{\alpha}}, \Psi_{\boldsymbol{\beta}}\rangle = \delta_{\boldsymbol{\alpha\beta}},
\end{equation}
with $\delta_{\boldsymbol{\alpha\beta}} = \prod\limits_{i=1}^M \delta_{\alpha_i\beta_i}$.
For practical purposes, the infinite series in Eq.~\eqref{PCE_Total} is truncated through a suitable truncation scheme, typically the maximum allowed polynomial degree $p$ \citep{xiu2002wiener}. For a review of more advanced basis truncation/construction schemes, the reader is redirected to \citet{luethen2021automatic}.
To calculate the set of expansion coefficients $a_{\ve{\alpha}} = \langle \cm(\vx), \Psi_{\ve{\alpha}}\rangle $, sparse regression techniques, also known as compressive sensing, have been demonstrated to be highly efficient in the context of surrogate modeling \citep{torre2019data,luethen2021sparse,luethen2021automatic}, and are adopted in this paper.
Once $a_{\ve{\alpha}}$ is calculated from a finite data set, the surrogate forward modelling predictor $\tilde{M}(\vx)$ can be calculated at a negligible cost, by direct evaluation of Eq.~\eqref{PCE_Total}. The accuracy of the surrogate model can be learned by comparing the performances of $\mathcal{M}$ and $\mathcal{\tilde{M}}$ on a validation set if available, or by using cross-validation techniques \citep{blatman2011adaptive,UQdoc_14_104}.

\subsection{Modeling-error aware Likelihood function}
A PCE surrogate model can be used to faithfully predict physical quantities such as $\vy$ in Eq. \eqref{forwardreduced}. However, since we rely on an imperfect forward model operating on a truncated domain it is important to incorporate  the corresponding error in the inversion process by using an appropriate likelihood function.
The likelihood function discussed in Section \ref{effective} is assumed to be Gaussian, and can be designed to ignore or incorporate the modelling error associated with the PCE and the PCA truncation.
We account for the modeling error by considering in the inversion process a covariance operator $\boldsymbol{C}_D$ given by the sum of the covariance matrices describing data uncertainty, indicated in the following as $\boldsymbol{C}_d$, and a modelling term $\boldsymbol{C}_{Tapp}$.
The covariance operator associated with the modelling error is defined in terms of a misfit matrix $\boldsymbol{D}_e$. For an unbiased model, $\boldsymbol{D}_e$  consists of $T$ realizations of the difference between the exact and the approximate forward responses. In case of biased models, the bias - indicated in the following as ${\dd}$ is added to the approximate responses \citep{hansen2014accounting,madsen2018estimation}. 
The $k^{th}$ column of $\boldsymbol{D}_e$ is a $n$-dimensional vector corresponding to the single $k^{th}$ realization of the modelling error, where $n$ is the dimension of the output domain, thus, $\boldsymbol{D}_e$ is a $n \times T$ matrix.
Given the definition above, the covariance operator $\boldsymbol{C}_D$ is defined as follows:
\begin{equation}
\label{CovMat}
    \boldsymbol{C}_{D} = \boldsymbol{C}_d + \boldsymbol{C}_{Tapp},
\end{equation}
where $\boldsymbol{C}_{Tapp}$ is defined as the $n \times n$ matrix
\begin{equation}
\boldsymbol{C}_{Tapp}    = \frac{1}{N} \boldsymbol{D}_e \boldsymbol{D}_e^T.
\end{equation}
%Note that under the assumption of i.i.d. Gaussian variables considered here (see Appendix A for more details), $\mathbf{C}_d = \sigma^2 \mathbf{I}$, where $\mathbf{I}$ is the $n \times n$ identity matrix.
 The likelihood incorporating the modelling error is then expressed as
      \begin{equation} 
      \label{likely}
   \mathcal{L}(\vx)=   \left( \frac{1}{2 \pi} \right)^{n / 2} |\boldsymbol{C_D}|^{-1/2} \mbox{exp} 
           \left[ -\frac{1}{2} (\mathcal{\tilde{M}}{(\vx)} + {\dd} - {\vy} )^T {\boldsymbol{C_D}}^{-1} (\mathcal{\tilde{M}}{(\vx)} + {\dd} - {\vy}) \right] \,.
\end{equation}

%The data used to define the matrix $\boldsymbol{D}_e$ does not necessarily need to be part of the training set. However in general cases all available information should be used to calibrate the model. 
In practice, it is convenient that the training set itself is used to synthesise the matrix $\boldsymbol{D}_e$.
The computation of the $k^{th}$ column of the matrix $\boldsymbol{D}_e$ is defined in terms of the output of the $k^{th}$ element of the training set and a PCE approximation. This PCE estimate is based on cross-validation in which the results of the surrogate trained on the all the training input except the $k^{th}$ one are used.%, that is, the corresponding leave-one-out set.
%In our numerical tests we used $50$, $500$ and $900$ training sets each consisting of $35$ output, which resulted in $50 \times 35$, $500 \times 35$, and $900 \times 35$ matrices for $\boldsymbol{D}_e$, respectively.

\section{Application to GPR crosshole tomography}
\begin{figure} 
  \centering
   \includegraphics[width=1\textwidth]{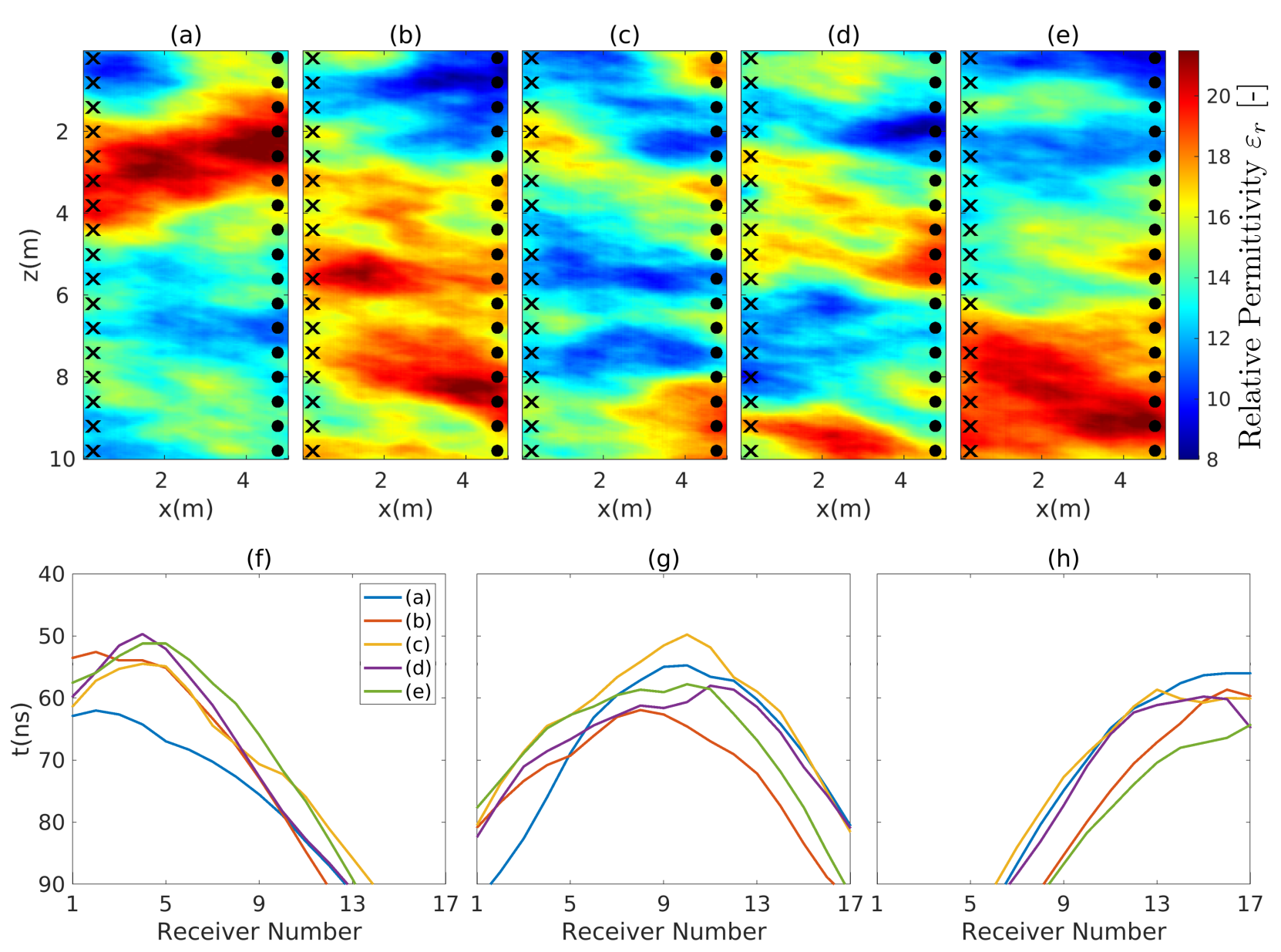}
\caption{(a-e) Representative realizations of the generative model and (f-h) exemplary corresponding travel-time gathers. Crosses and circles stand for sources and receivers, respectively.}
\label{Img_001}
 \end{figure}

We apply now the proposed methodology to a GPR crosshole tomography problem. We consider for simplicity a GPR tomographic problem involving variations in permittivity, while assuming a constant and known electrical conductivity. We use the recording configuration displayed in Fig. \ref{Img_001}(a) with 17 evenly spaced sources and receivers located in two vertically-oriented boreholes. The distance between the boreholes is 4.6 m, while the spacing between sources/receivers is 0.6 m. Each source is characterized by a $100$ MHz Blackman–Harris  pulse that is fired separately and the propagating signals are recorded at all receivers, such that a total of 289 travel-times are collected. We employ a 2D FDTD solver to simulate noise-free propagation in the transverse-electrical (TE) mode \citep{irving2006numerical}. Perfectly matched layers (PML) surrounding the propagation domain is used to prevent spurious reflections from contaminating the data, while appropriate space-time grids are employed to avoid instability and dispersion artefacts. Travel-times are picked automatically based on a threshold associated with the relative maximum amplitude of each source-receiver pair. We rely on a generative model representing prior knowledge to create a set of relative permittivity distributions $\epsilon_r$ typical of a near-surface geophysical setting. We consider $5$ m wide, $10$ m deep  random multi-Gaussian field realizations created using a pre-defined covariance structure based on the 2-D Matérn geostatistical model \citep{dietrich1997fast,laloy2015probabilistic}. The models are generated using six parameters: the mean (15), the standard 
deviation of $\epsilon_r$ (2.45), the anisotropy angle (85 degrees), the anisotropy ratio (0.3), the integral scale of the major axis (10 m) and a shape parameter (1.15). 
Figures \ref{Img_001}(a-e) display representative samples of $\epsilon_r$-distributions, together with a few representative gathers in Figs. \ref{Img_001}(f-h). 
Given the permittivity range and the frequency content of the source wavelet, a spatial grid with $dx=dz=4$ cm is used to avoid numerical dispersion and instability. Each generated permittivity field corresponds to a $125\times250$ dimensional space imposed by the FDTD grid. 
%In the next section we show how the map between input and output can be faithfully represented using a PCA-parameterized input domain characterized by only few tens parameters.
\label{Generative_Model}
\subsection{Parametrization of input and output domains}
\label{parametrization}
\begin{figure} 
  \centering
   \includegraphics[width=1\textwidth]{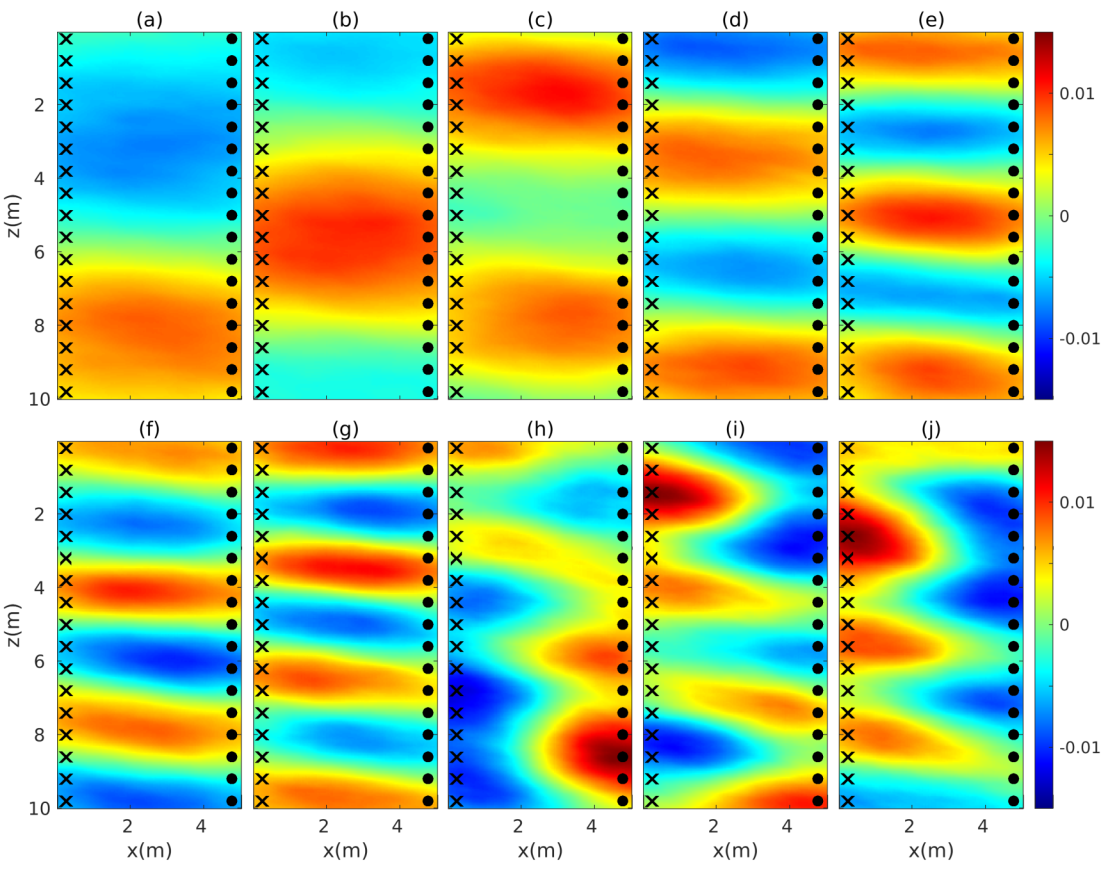}
\caption{(a-j) The first 10 principal components in the input domain. Crosses and circles stand for sources and receivers, respectively.}.
\label{Img_002}
 \end{figure}

Following a similar strategy to \citet{laloy2013efficient}, we use the generative model to learn a prior-knowledge-based parametrization of the input domain. We create a total of $1000$ random realizations from the prior (see Fig. \ref{Img_002}). This number of realizations allowed inferring an appropriate sample mean and principal components of the underlying prior distribution, which led to convergence in terms of input domain  root-mean-square error (in the following, RMSE) and Structural SIMilarity of the projections of distributions on subsets of principal components \citep{wang2004image}. Rather than considering the permittivity realizations to define the principal components, we take at each grid point the square root of each permittivity sample. This is done to simplify the PCE learning as the relationship between $\sqrt{\epsilon_r}$ and travel-times is perfectly linear in case of constant velocity. However for readability purposes comparison of projections and inversion results will be always referred to relative permittivity distributions.
%As also observed in \cite{giannakis2021fractal}
As expected, low wavenumber features are mainly associated with the first principal components, whereas small-scale details can be identified in higher modes (Fig. \ref{Img_002}). %The effective dimensionality of the forward problem is assessed by analysing the convergence to the reference solution in the output domain as a function of the number of principal components used to represent the input.
\begin{figure} 
  \centering
   \includegraphics[width=1\textwidth]{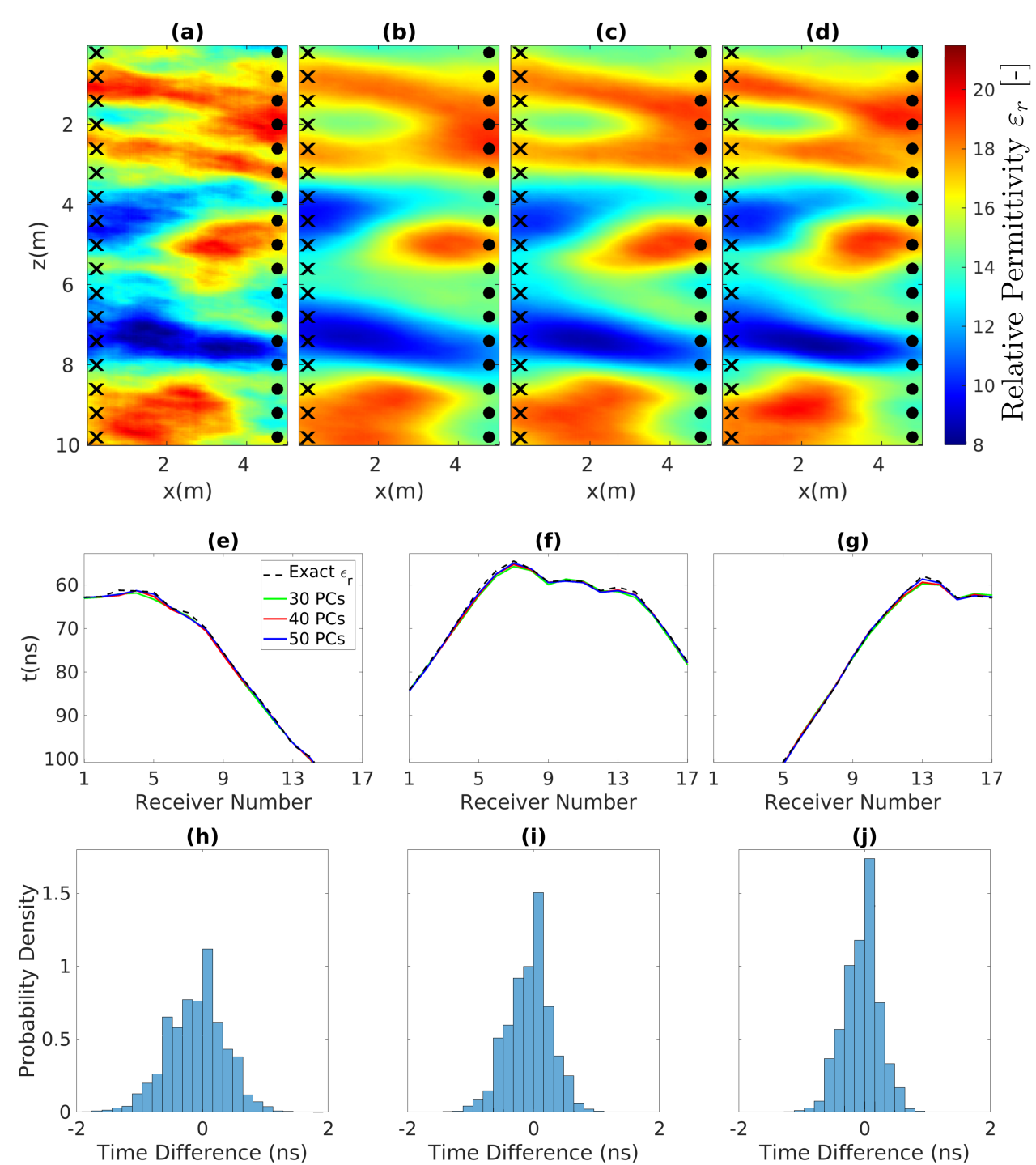}
\caption{(a) The reference relative permittivity distribution used in the numerical inversion experiment and its representation using the first (b) 30, (c) 40 and (d) 50 principal components, and (e-g) exemplary corresponding travel-time gathers and (h-j) histograms of travel-time residuals based on FDTD forward simulations on the true field and reconstructions based on principal components. Crosses and circles stand for sources and receivers, respectively}.
\label{Img_003}
 \end{figure}
In Fig. \ref{Img_003}(a), a permittivity distribution is shown next to the approximate representations (Fig. \ref{Img_003}(b-d)) obtained by projecting it on a moderate number of principal components. The remaining terms are ignored by setting their coefficients to $0$, which corresponds to the mean value of their respective component. 
 Table \ref{table:0} summarizes the convergence in the relative permettivity domain of the various projections in terms of RMSE and Structural SIMilarity.
\begin{table}
\begin{center}
\caption{Assessment of the convergence of projections of permettivity realizations on a number of principal components in terms of input RMSE and SSIM. Note that the mean of the RMSE for a random realization of the generative model is $\sim 15.2$.}
\begin{tabular}{||c c c||} 
 \hline
 PCA  & RMSE $\epsilon_r$ & SSIM $\epsilon_r$ \\ [0.5ex] 
 \hline\hline
 30 & 0.88 & 0.70 \\ 
 \hline
 40 & 0.80 & 0.72 \\
 \hline
 50 & 0.69 & 0.75 \\
 \hline
 
\end{tabular}
\label{table:0}
\end{center}
\end{table}

The similarity between Fig. \ref{Img_003}(a) and Figs. \ref{Img_003}(b-d) refer to proximity in the input domain, and does not guarantee that the map between input and output can be faithfully represented by operating on the reduced representations. The fidelity of such a reduced representation must, thus, also be assessed by considering the performance in the output domain. 
This aspect is is highlighted in Figs. \ref{Img_003}(e-g),  showing representative travel-time gathers associated with the distributions in Fig. \ref{Img_003}(a-d). 
Each approximate distribution seems to provide similar responses as those generated by the reference input. To quantify the agreement of the various reduced parametrizations, we consider 100 realizations of the generative model, and compute the resulting histograms of the travel-time residuals. The output domain root-mean-square error (in the following, rmse) of the misfit between the data associated with the original distribution and its projections on $30$, $40$ and $50$ principal components are $0.46$, $0.34$ and $0.28$ ns, respectively, which are significantly smaller than the expected level of GPR data noise of $1$ ns for $100$ MHz data \citep{arcone1998ground}.
Based on this analysis, we can claim that the problem can be faithfully approximated involving forward simulations of fields with 30-50 of prior-knowledge-informed parameters.

\begin{figure} 
  \centering
   \includegraphics[width=1\textwidth]{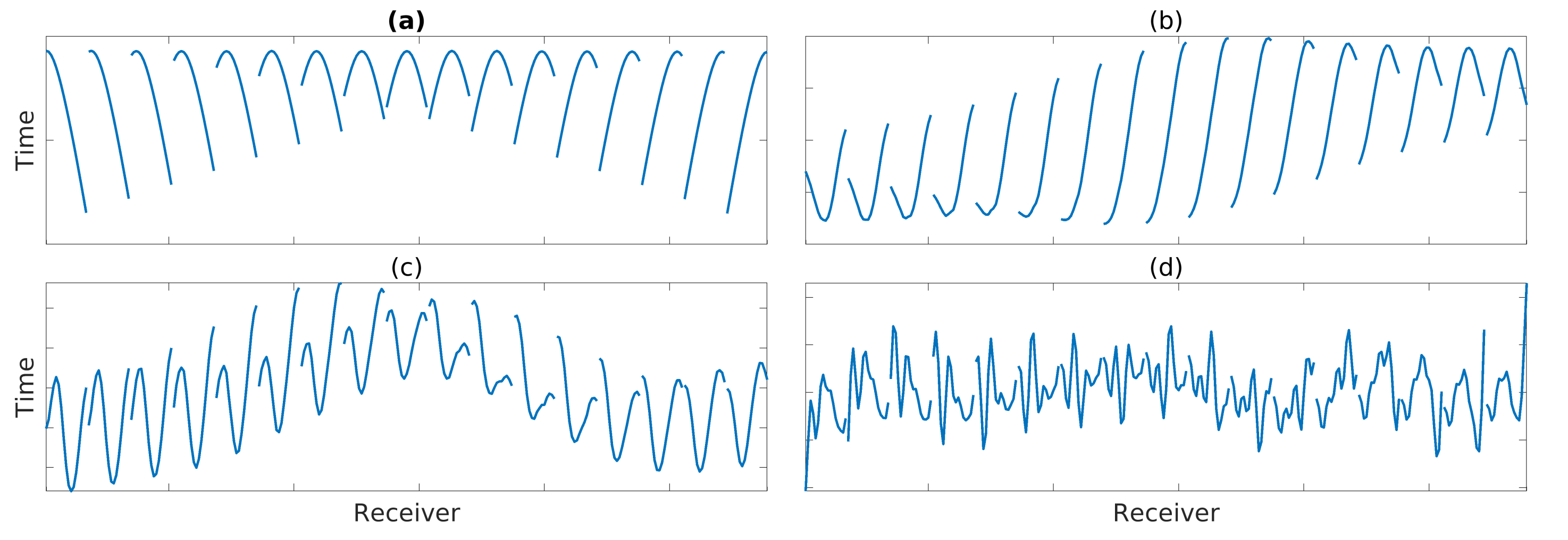}
\caption{(a) Sample mean and (b-d)  representitative principal components in the output domain}.
\label{Img_004}
 \end{figure}

\begin{figure} 
  \centering
   \includegraphics[width=1\textwidth]{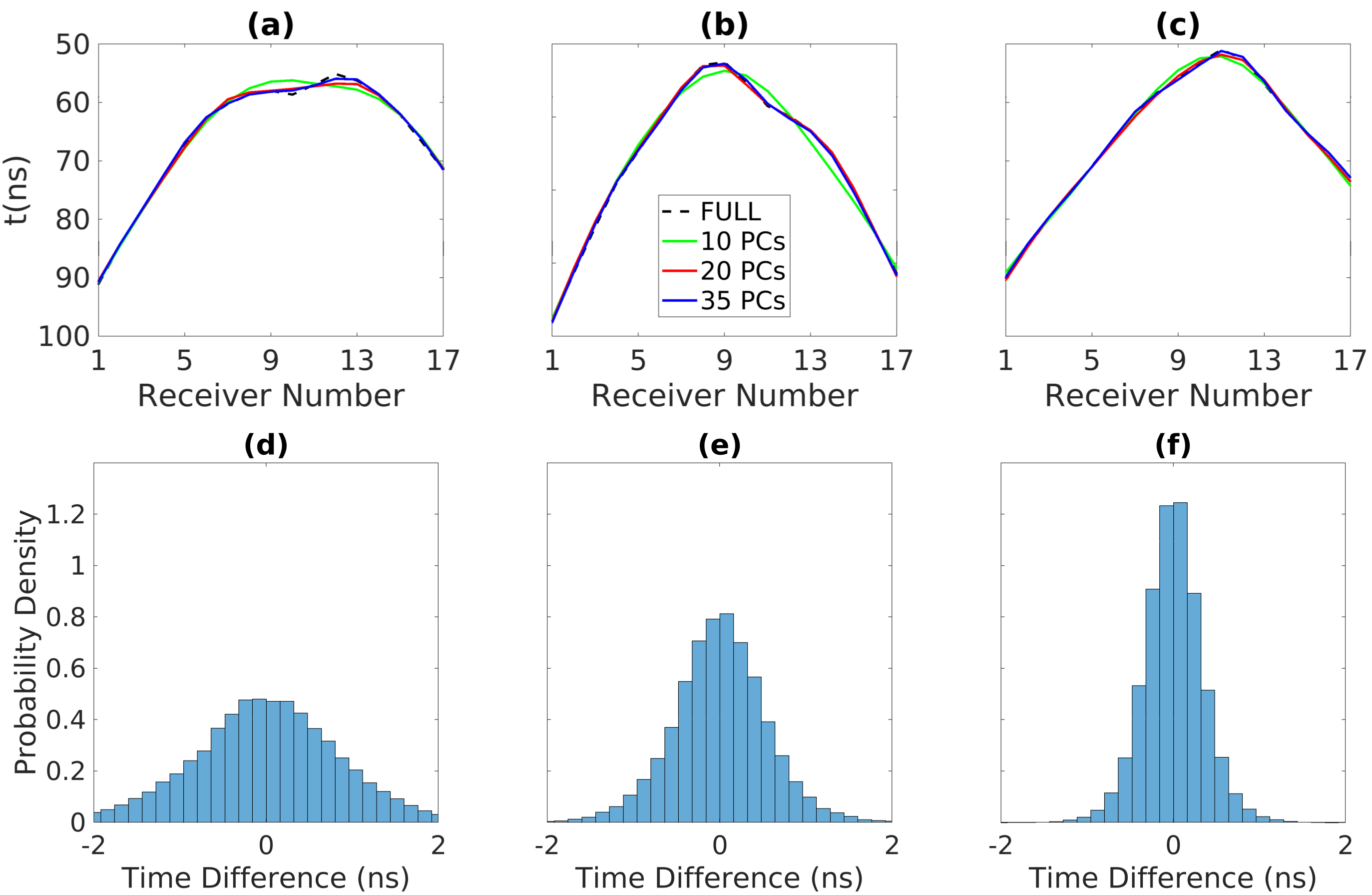}
\caption{Decomposition in terms of (a) 10, (b) 20 and (c) 35 principal components in the Gather output domain of three representative gathers associated with a realization of the generative model, and (d-f) corresponding histograms of the residuals. The computation of the reference data is carried out using the FDTD solver.}.
\label{Img_005}
 \end{figure}

Dimensionality reduction can also be exploited in the output domain to further reduce computational costs, storage and PCE training. 
We consider 1000 data sets corresponding to as many realizations of the generative model. We then compute via FDTD the 289 travel-times associated with the recording configuration discussed above.
Based on these gathers we create a training set from which principal components are retrieved. The original data can then be projected on a subset of the principal components and characterized by the corresponding coefficients.
The sample mean and representative principal components are shown in Fig. \ref{Img_004}, while projections of selected source gathers are shown in Figs. \ref{Img_005}(a-c).
The agreement with the original data (black dashed line) and the projection on the first $10$ (Fig. \ref{Img_005}(a)) components transformed back into the original output domain is rather poor, but improves significantly when $20$ (Fig. \ref{Img_005}(b)) and $35$ (Fig. \ref{Img_005}(c)) components are used. Corresponding histograms of the travel-time residuals in Figs. \ref{Img_005}(d-f) provide more quantitative insights concerning the performances of the compression. When only $10$ principal components are employed the rmse is $0.94$ ns. The fitting improves when $20$ and $35$ principal components are used, as the rmse reduces to $0.55$ and $0.34 $ ns, respectively.
%Note that PCA is defined via an orthogonal linear transformation, and Principal Components form an orthonormal basis. In our case, each new output $Y_k$ is defined as a linear combination of data, namely: $Y_j = \sum_{i=1}^{N} a_ij W_i -$, where $W_i$ is associated with a specific travel-time and $N$ is the total number of source/receiver combinations. We assume here that $W_i, i \in 1:N$  constitute a set of independent identically distributed Gaussian variables characterized by $\sigma^2=1^{-18}ns^2$.
%The projection on principal components is defined via scalar product on an orthonormal basis. 
%In our case, the new output values are defined as a linear combination of travel-times, which are assumed to be Gaussian independent and identically distributed variables characterized by $\sigma^2=1ns^2$. 
%The same independence and variance is shared by the new and original set of variables due to the orthonormal properties of the operator implementing the principal components decomposition (see Appendix for a derivation).

Despite the massive dimensionality reduction offered by PCA-decomposition, an FDTD-based Monte Carlo inversion would be extremely computationally demanding to carry out. 
To circumvent this and given the moderate number of input parameters, we replace the forward simulations with a PCE-based surrogate model, as implemented in the Matlab Package UQlab \citep{marelli2014uqlab,UQdoc_14_104}. 
Note that the implementation of a PCE requires a prior probability density function to build a corresponding set of orthonormal polynomials and the computation of a design set to train the algorithm.
We follow the approach described in \citet{torre2019data}, by assuming statistical independence between the modes, and use kernel smoothing to estimate data-driven marginal PDFs.
In particular, we re-use the realizations of the generative prior model to calculate the orthonormal polyomial basis.

Because PCE is designed for models with scalar outputs, the main benefit of compressing the multivariate forward model output is to reduce the number of PCEs that need to be trained and evaluated. 
Using compression in the output then reduces the CPU time required to train the PCs, while additionally reducing  the overall memory fingerprint of the MCMC process.

The proposed implementation of PCE relies on peculiar representations of both input and output domains. To train the PCE via FDTD modelling we use a pixel-based parametrization of $\sqrt{\epsilon_r}$. However, unless stated differently, to offer easy to interpret images we present relative $\epsilon_r$ distributions in the pixel-based domains rather then the corresponding projections on principal components. A similar distinction is important for output representation as well. FDTD modelling generates travel-time gathers which are projected on  principal components before being used in the PCE process and inversion.
The PCE output is defined in a domain that does not offer an immediate physical interpretation, but carry important statistical properties (see Appendix 1). To comment our results, in the following we will then use both the PCE output domain (in the following, 'PCE output') and the back-projection to the FDTD gather domain (in the following, 'Gather output').

\subsection{PCE performance}
\label{PCEperformance}
\begin{figure} 
  \centering
   \includegraphics[width=1\textwidth]{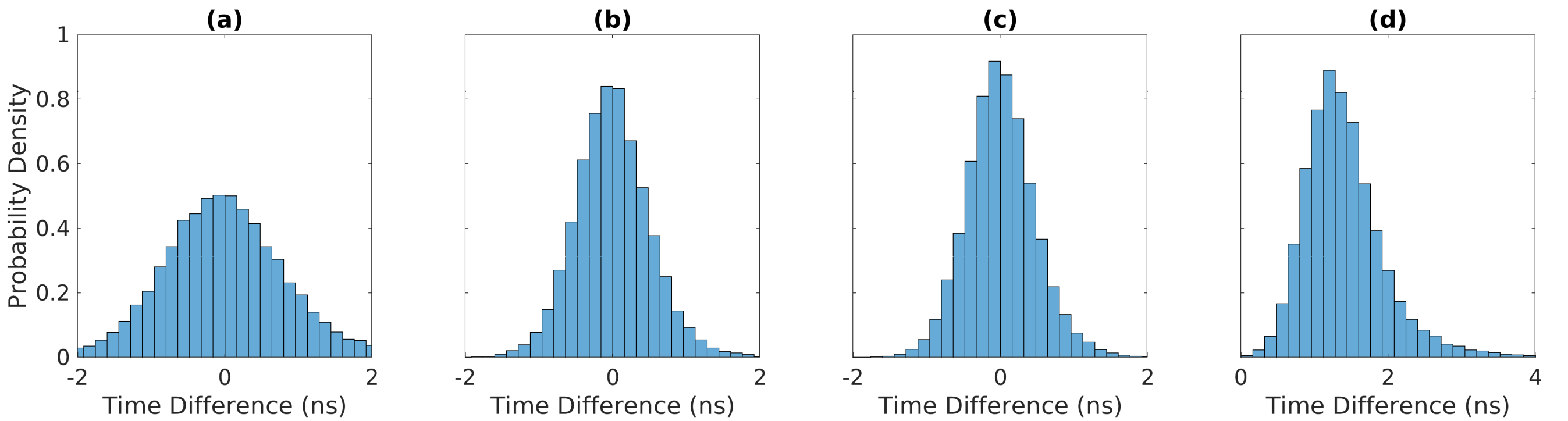}
\caption{(a) Histograms of the model errors with respect to the PCE prediction when using 50 and 35 principal components in the input and output domain, respectively, and 50 training sets for the expansion. (b) and (c) are constructed with the same parameters but with 500 and 900 training sets, respectively. (d) Histograms of the model errors associated with the straight-line-approximation. The computation of the reference data is carried out using the FDTD solver.}
\label{Img_007}
 \end{figure}

Based on considerations of the expected noise level in field data and the performances of the proposed dimensionality reduction in the GPR problems discussed above, we perform PCE using $50$ principal components in the input and and $35$ in the PCE output domain. With these settings, we analyze the PCE performance as a function of the training set size. 
In this study, the output is trained to predict the projection coefficients of travel-times corresponding to all source-receiver pairs using a size of $\vx$ ($M$ in Eq.~\eqref{PCE_Total}) that is $50$. We set $p$, the maximum allowed polynomial degree, to $5$. %Local training is a requisite to ensure high-performance surrogate-modeling from moderate training sets and a rather low maximal polynomial degree. In addition, stability is offered by using sparse adaptive PCE in order to limit the number of non-zero entries in the inferred polynomials. 
Histograms of travel-time residuals between the reference travel-time data and PCE results obtained using $50$ principal components in the input domain and $50$ (Fig. \ref{Img_007}(a)), $500$ (Fig. \ref{Img_007}(b)) and $900$ (Fig. \ref{Img_007}(c)) training sets. 
%Note that the reference data are computed by the FDTD solver on uncompressed input data, and this can lead to spectral leakage that affects the estimates of the coefficients of PCEs based on compressed input data if the higher truncated components are not in the null-space \citep{trampert1996model,linde2013distributed}. Based on the histograms in Figs. \ref{Img_003}(j) and \ref{Img_006}(c) we assume that projection of the input on $50$ principal components accurately predict actual travel-times as generated by the uncompressed input. We therefore assume the leakage to be minor when $50$ principal components are used.
When only $50$ training sets are used, the rmse in the Gather output is $0.86$ ns (Fig. \ref{Img_007}(a)). With an increasing size of the training set, the PCE performances improve significantly, with rmse of the misfit in the Gather output reducing to $0.51$ and $0.47$ when $500$ (Fig. \ref{Img_007}(b)) and $900$ (Fig. \ref{Img_007}(c)) data sets are employed, respectively. %hese latter numbers are comparable to what is achieved when operating via FDTD on the reduced input domain (see Fig. \ref{Img_006}(c)).
Note that regardless of the number of training data sets, the PCE results are unbiased and quite closely resemble Gaussian distributions.

By using only $50$ principal components to characterize the input domain for the PCE predictions, we assume that higher-order principal components belong to the null space of the forward problem. 
While not strictly valid, this assumption is partially justified by the histograms in Figs. \ref{Img_003}(j) and \ref{Img_005}(f). 

For comparison purposes, we now show the corresponding histogram of the travel-time residuals obtained when using the straight-line-approximation (Fig. \ref{Img_007}(d)). 
As expected, the straight-line-approximation always overestimates the exact travel-times, and has an rmse of $1.53$ ns. By removing the mean value of the residuals (bias), the rmse reduces to $0.52$ ns, which is still higher than for the PCE results using a training set size of 500 and above. In contrast to the PCE results, the straight-line-approximation histogram is clearly asymmetrical and, hence, non-Gaussian, which makes the likelihood in Eq. \eqref{likely} inaccurate as it is based on a Gaussian assumption.
The computational burden involved in constructing the PCE is comparable to what is required to account for modelling error for  the straight-line-approximation (i.e., the evaluation of FDTD training sets) and its evaluation cost is negligible with respect to the computation time of the reference FDTD solver. 
Most of the CPU time required to learn the PCE is due to the calculation of the training set, while the evaluation on a given input is several orders of magnitude faster than the corresponding FDTD solution. For a single forward run, PCE is about $1000$ times faster than FDTD. However, in the MCMC process multiple forward runs can be performed at once (e.g. when multiple chains are updated simultaneously), and the massive parallelization capability provided by PCE is, therefore, particularly advantageous. For example, evaluation of the PCE on $10000$ input models at the same time costs $\sim+20\%$ than what is required by the evaluation of the PCE on a single input. 

\subsection{Inversion results}

\begin{figure} 
  \centering
   \includegraphics[width=.9\textwidth]{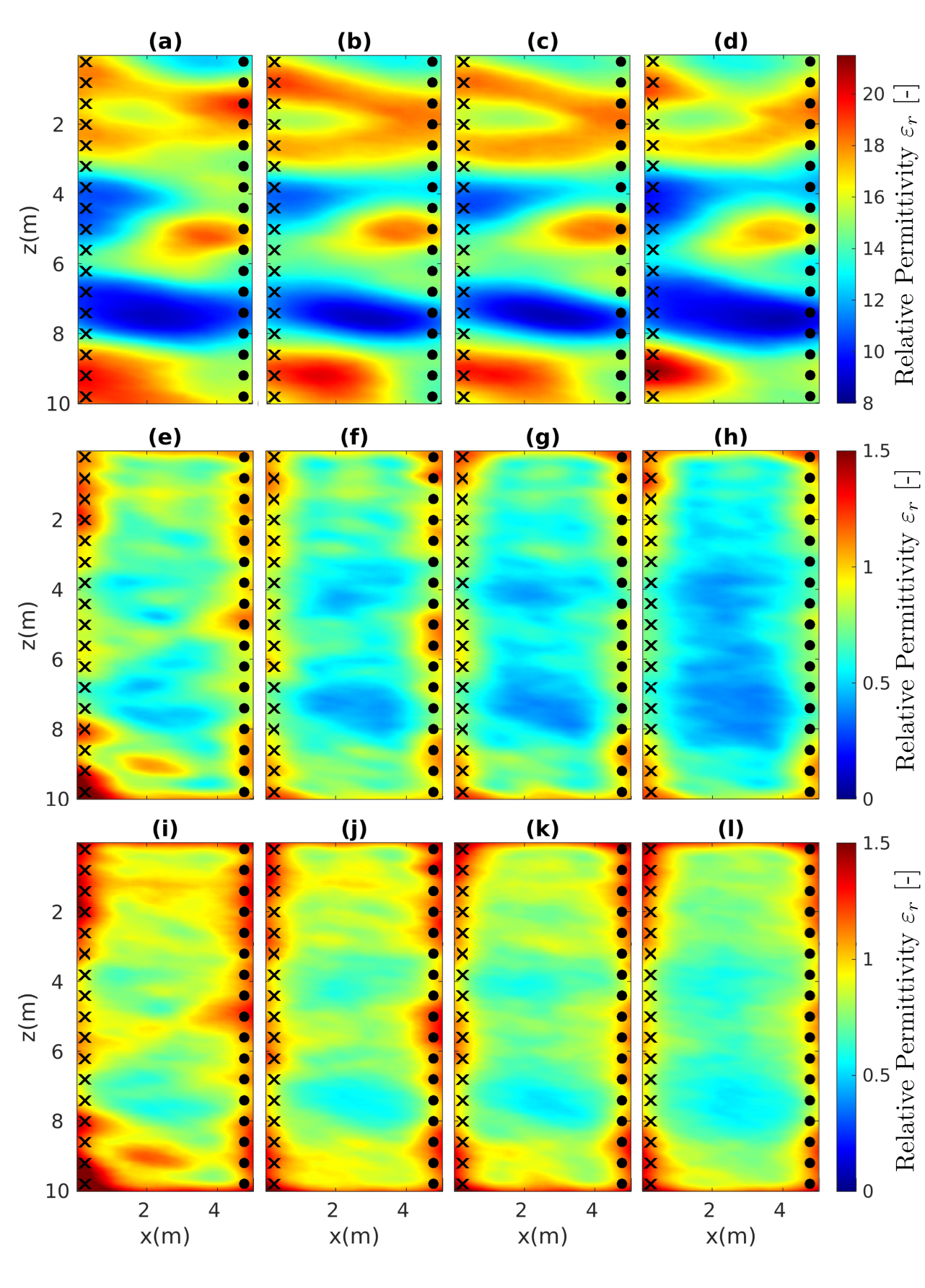}
\caption{MCMC-based posterior mean obtained while ignoring modeling errors in the likelihood function with PCE based on (a) 50, (b) 500 and (c) 900 training sets and (d) using a straight-line-approximation solver. The PCE operate between reduced domains defined in terms of 50 and 35 principal components associated with $\sqrt{\epsilon_r}$ and travel-time gathers, respectively. Results are shown as $\epsilon_r$ distribution for readability purposes. Corresponding posterior standard deviations (e-h) without considering the null-space and (i-l) considering the null-space. Crosses and circles stand for sources and receivers, respectively}
\label{Img_009a}
 \end{figure}

\begin{figure} 
  \centering
   \includegraphics[width=1\textwidth]{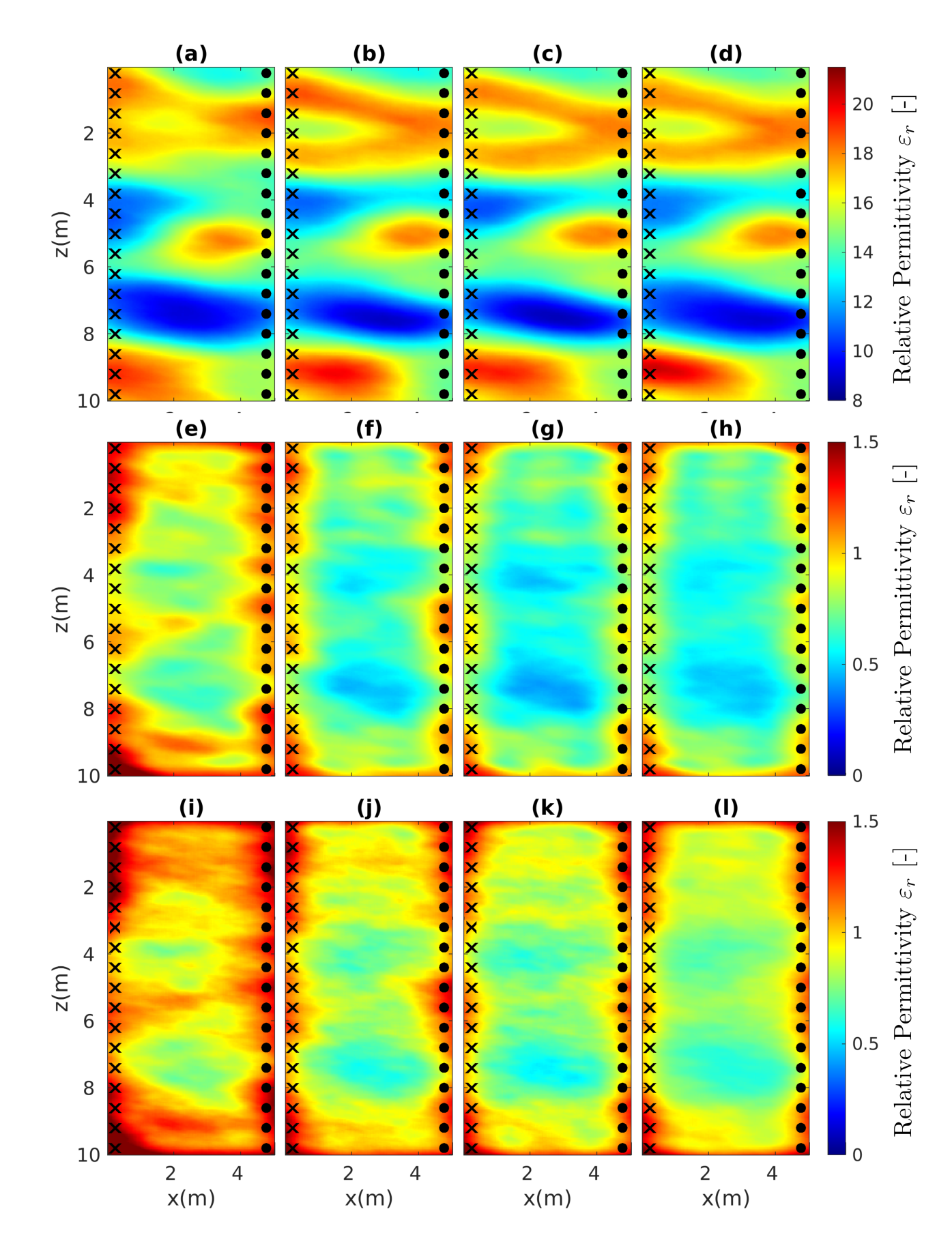}
\caption{MCMC-based posterior mean obtained while incorporating modeling errors in the likelihood function with PCE based on (a) 50, (b) 500 and (c) 900 training sets and (d) using a straight-line-approximation solver, corresponding posterior standard deviations (e-h) without considering the null-space and (i-l) considering the null-space. Crosses and circles stand for sources and receivers, respectively}
\label{Img_009b}
 \end{figure}

In section \ref{PCEperformance}, we have shown how a PCE can approximate a forward FDTD solver in the computation of travel-times in a realistic geophysical scenario operating on a reduced input domain. Importantly, its performance using 500 training sets lead to modeling errors that are much less than the expected data noise and it is notably better than straight-line approximations that are biased, have skewed error statistics with larger variance. %We have also seen that accurate PCEs can be obtained using moderate training sets. 
Here, we explore the potential of PCEs based on $50$ PCs in the input domain (obtained using $50$, $500$ and $900$ training sets, respectively) and a straight-line solver in performing probabilistic MCMC inversion. 
The entire inversion process was carried using the UQLab Matlab package framework \citep{UQdoc_14_113}.
We invert for the permittivity distribution shown in Fig. \ref{Img_003}(a), which is used without model compression to generate travel-times using the reference FDTD solver. Uncorrelated Gaussian noise characterized by $\sigma^2=1 ns^2$ is added to the FDTD data. Compression in the output domain is then applied, and the resulting data are inverted using the PCE surrogate model.
We use a Metropolis-Hastings algorithm, and run $15$ Markov chains in parallel for $1.5\times10^5$ iterations. Based on the visual inspection of the chains, we discard the first $1000$ iterations as burn in for the evaluation of mean and variance values of the final results. The scaling factor of the proposal distribution is tuned so that an acceptance rate of about $30\%$ is achieved for each experiment.  
Inversion results in terms of mean and standard deviations ignoring the model error (i.e. setting both $\boldsymbol{C}_{Tapp}=0$ and ${\dd}=0$in Eqs. \eqref{CovMat} and \eqref{likely}, respectively) are shown in Fig. \ref{Img_009a}.
The PCE and Gather output rmse associated with the PCE-simulations of the posterior mean results in Figs. \ref{Img_009a}(a-c) are $2.5$ and $1.44$ ns for a training set of $50$, $1.24$ and $1.18$ ns for a training set of $500$ and $1.10$ and $1.19$ ns for a training set of $900$, respectively. The slightly worse result associated with the Gather output for the PCE trained with $900$ data sets is strictly speaking not indicative of a worst inversion, since the cost function is defined in terms of PCE output data. The rmse associated with the posterior mean obtained with the straight-line-approximation (Fig. \ref{Img_009a}(d)) are significantly higher at $4.70$ and $2.05 ns$ in PCE and Gather output, respectively. 
Figures \ref{Img_009a}(e-g) show the corresponding standard deviations of the posterior distributions without considering the variability in the truncated PCA coefficients (the assumed null-space). High standard deviations are mainly located along the boreholes and the top and the bottom of the model due to the limited sensitivity of the data with respect to those areas. However, an accurate comparison of posterior uncertainty in the input domain can only be carried out if we consider elements from the null space. The corresponding standard deviation distributions obtained when adding the truncated PCA components (the assumed  null-space)(Fig. \ref{Img_009a}(i-l)) indicate an overall increase in posterior standard deviation ($\sim +20\%$), thus, demonstrating the importance of adding elements from the 'null space' to properly assess posterior uncertainty.
\begin{figure} 
  \centering
   \includegraphics[width=1\textwidth]{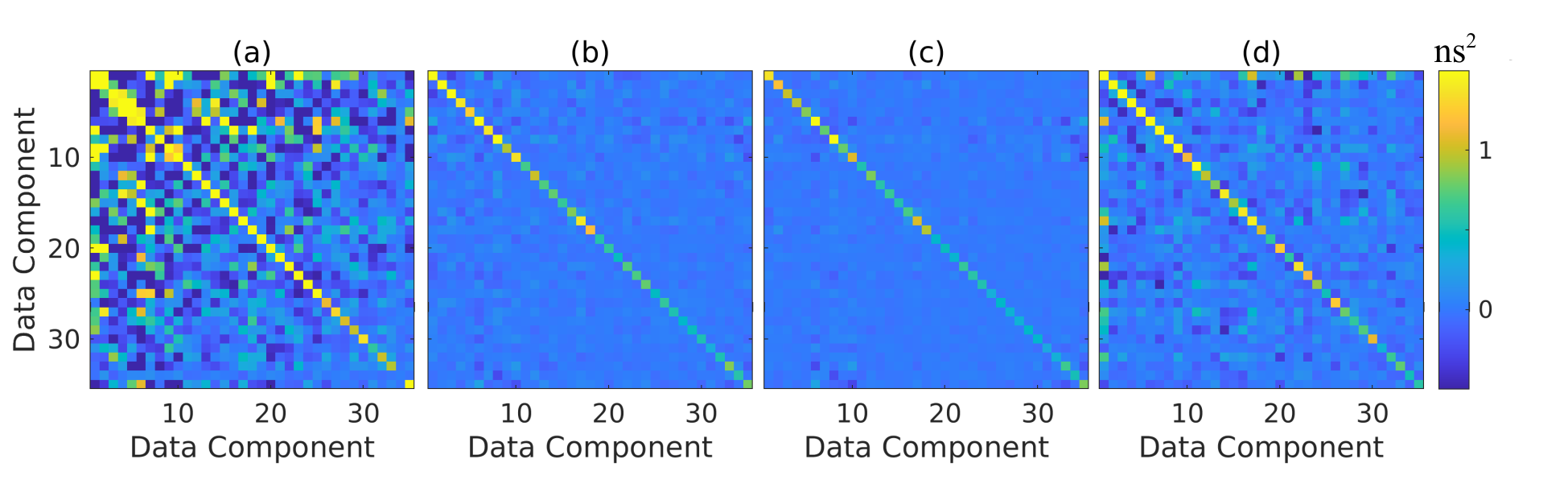}
\caption{Model error covariance matrices associated with PCE based on (a) 50, (b) 500 and (c) 900 training sets. (d) Model error covariance matrix associated with the straight-line solver}.
\label{Img_008}
 \end{figure}
We now run another set of inversions taking into account the modelling error to appreciate its effect on the corresponding solution. Fig. \ref{Img_008} shows covariance matrices $\boldsymbol{C}_{Tapp}$ of the modeling errors for the PCE and straight-line solver used in the previous example.  %When the modelling error is ignored, the results in Figs. \ref{Img_009a} for $50, 500$ and $900$ are obtained. 
The corresponding PCE and Gather output rmse values of the PCE- simulated data are $2.25$ and $1.41$ ns (Fig. \ref{Img_009b}(a)),$1.26$ and $1.16$ ns (Fig. \ref{Img_009b}(b)), $1.07$ and $1.17$ ns (Fig. \ref{Img_009b}(c)), rather similar to those of the previous example. However, the rmse for the straight-line inversion are significantly smaller at $1.54$ and  $1.26$ ns, for PCE and Gather output, respectively, (Fig. \ref{Img_009b}(d)), mainly as a consequence of the bias removal in the likelihood function. Standard deviation results are summarized in Fig. \ref{Img_009b}(e-l) and show a similar trend to what discussed about Fig. \ref{Img_009a}(e-l).
The results discussed above show that differences between PCE and straight-line-approximation-based inversion can be appreciated in terms of data misfit. 
Rather interestingly, ignoring or incorporating modelling error in PCE inversion seems to have little impact on the results in the input domain. The reason is most likely that the PCE is bias free and that the model error is significantly smaller than the observational errors.
The PCE-based posterior mean fields in Figs. \ref{Img_009a}(a-c) agree with the projection of the reference distribution on the first $50$ principal components (Fig. \ref{Img_003}(d)) similarly to what the results in Figs. \ref{Img_009b}(a-c)) do.
However, the two sets of solutions differ in terms of MCMC convergence, as assessed using the potential-scale reduction factor $\hat{R}$ considering the variance of the individual Markov chains with the variance of all the Markov chains merged together \citep{gelman1992inference}. Convergence is usually assumed if $R<1.2$ for all parameters. In our experiments, when modelling error is taken into account, all of the $50$ parameters have a value of $R$ smaller than $1.2$ for any choice of the training set dimension size. However, if modelling error is not taken into account, $R$ is larger than $1.2$ for at least some of the parameters, regardless of the size of the training set.
By contrast to the PCE results, the impact of incorporating model error in the inversion is very relevant when the (biased) straight-line solver is employed, as significantly better results are achieved when $\boldsymbol{C}_{Tapp}$ and ${\dd}$ are taken into account (compare the poor agreement of \ref{Img_009a}(d) and the good agreement of Fig. \ref{Img_009b} with respect to Fig. \ref{Img_003}(d)). As for the previous cases, convergence is achieved only if model error is taken into account.

\begin{figure} 
  \centering
   \includegraphics[width=1\textwidth]{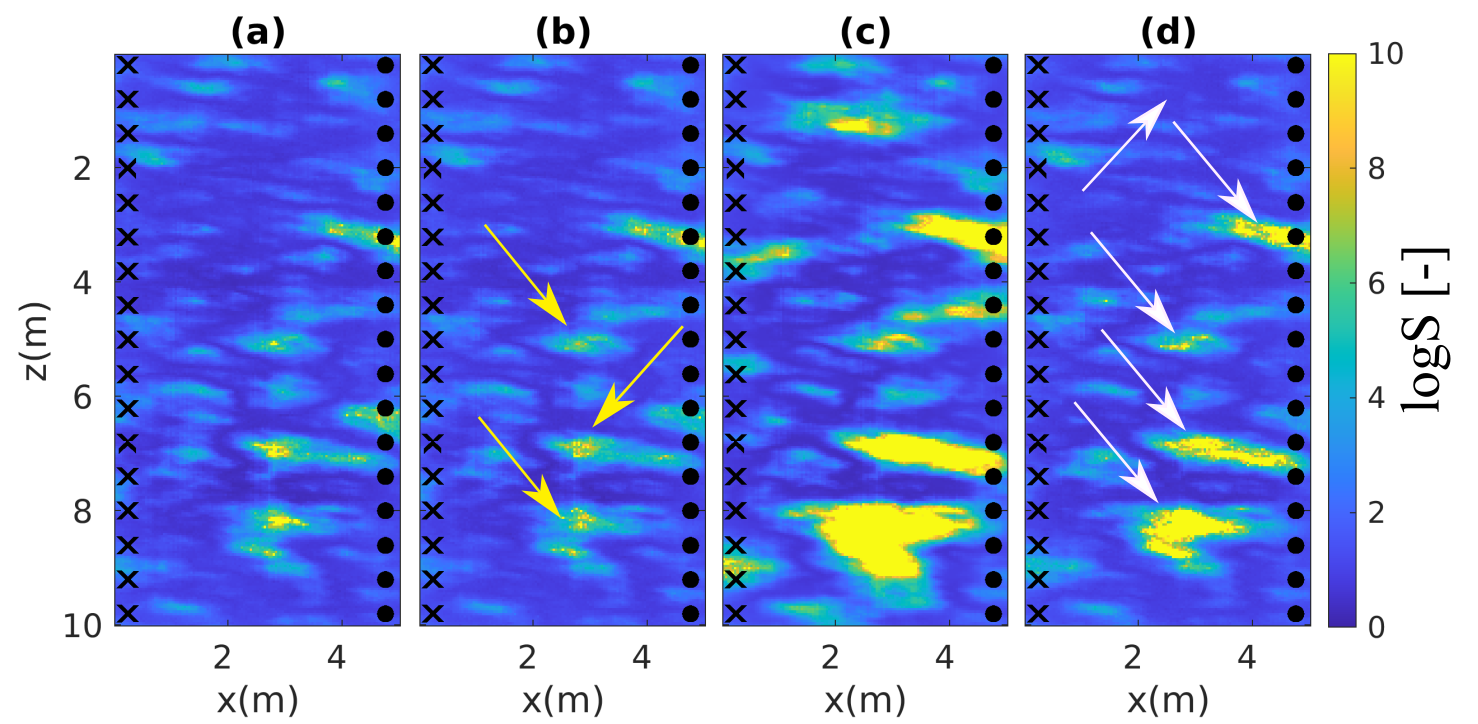}
\caption{Logarithmic score values with PCE (a) ignoring model error and (b) considering the model error in the likelihood function. Corresponding values for the straight-line-approximation (c) ignoring and (d) considering the model error in the likelihood function. Crosses and circles stand for sources and receivers, respectively}
\label{SCORING}
 \end{figure}

\begin{figure} 
  \centering
   \includegraphics[width=1\textwidth]{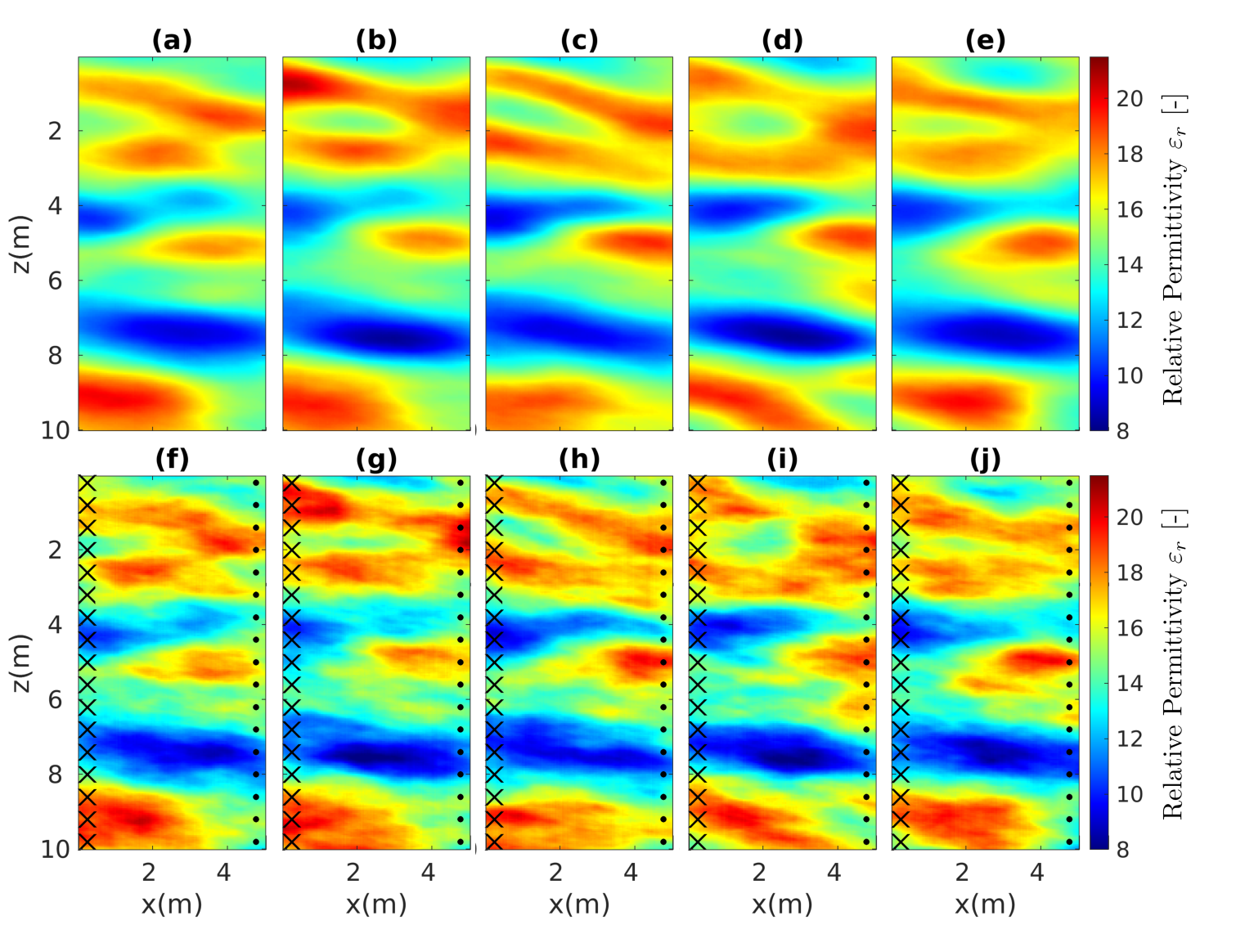}
\caption{Samples from the posterior distribution (a-e) ignoring the null-space and (f-j) corresponding realizations including random realizations from the null space. Crosses and circles stand for sources and receivers, respectively}
\label{Img_011}
 \end{figure}

\begin{figure} 
  \centering
   \includegraphics[width=1\textwidth]{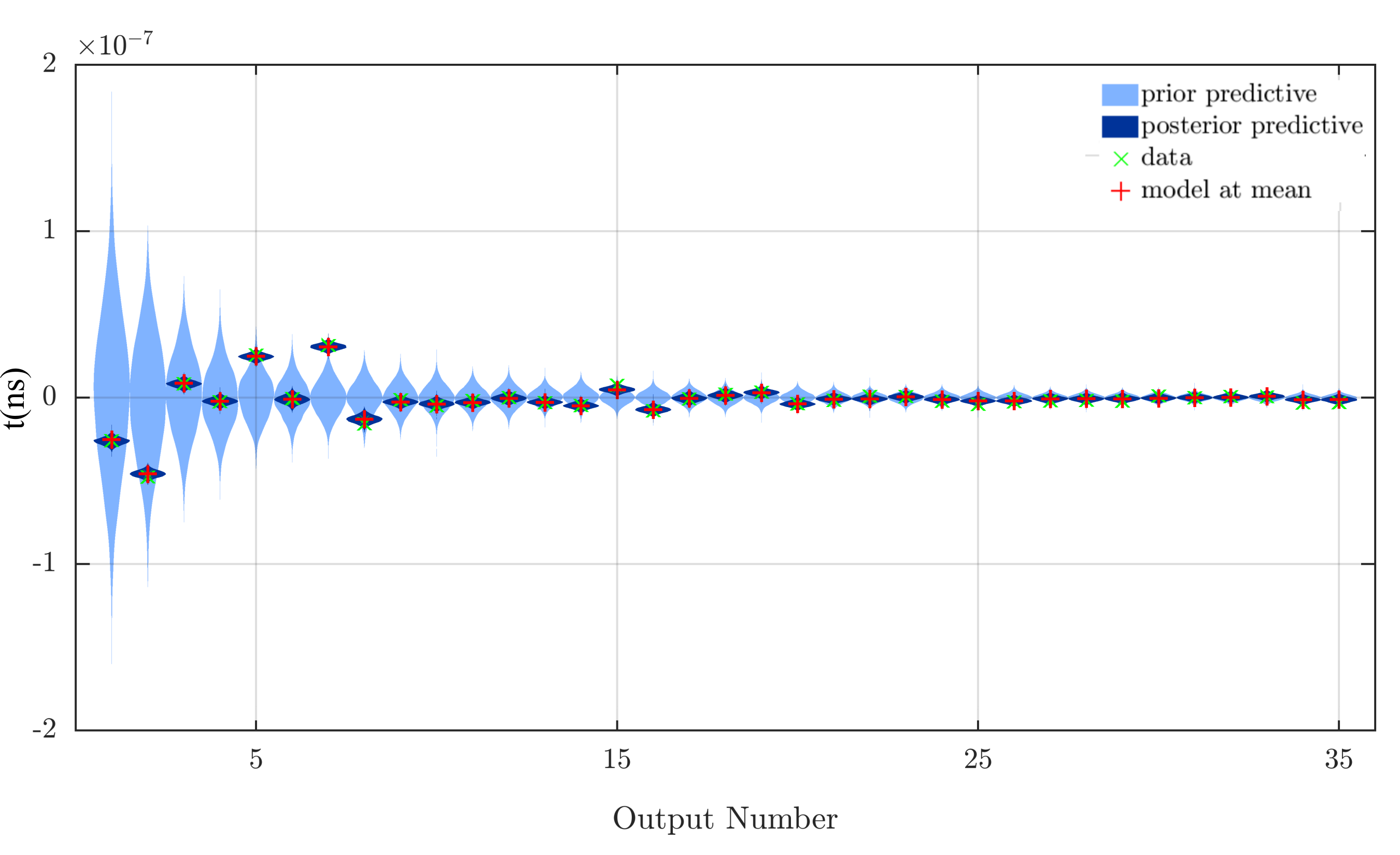}
\caption{Prior and posterior predictive distributions in the PCE output domain. Light and dark blue histograms in the output domain correspond to 10000 realization of prior and posterior distributions, respectively.
The green `$\times$' markers indicate the 'observed' data associated with the reference model, whereas the red `$+$' indicate the application of the PCE to the mean of the posterior.}.
\label{Img_012}
 \end{figure}

\begin{figure} 
  \centering
   \includegraphics[width=1\textwidth]{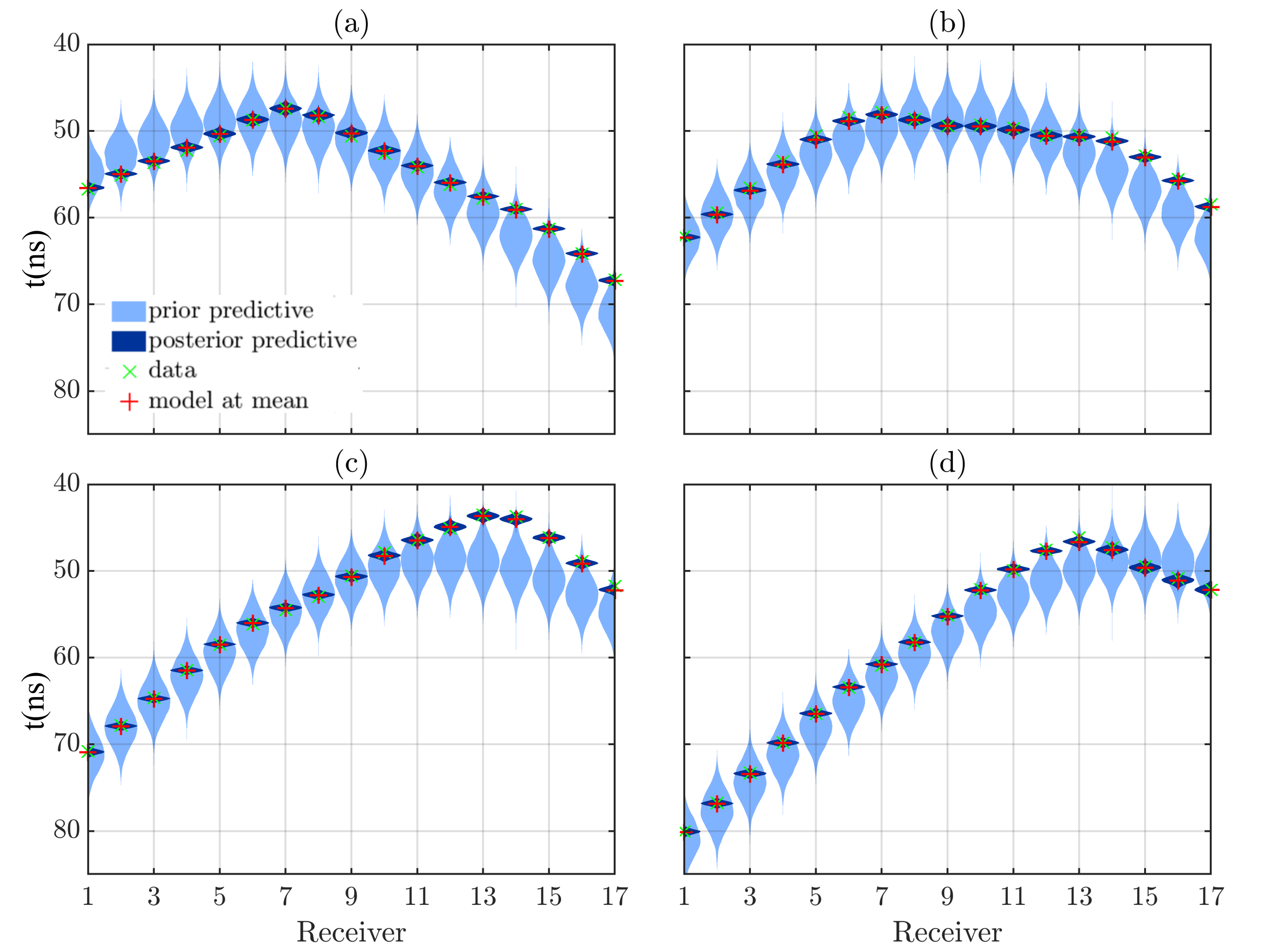}
\caption{(a-d) Exemplary prior and posterior predictive distributions in the travel-time Gathers output domain. Differently than the PCE output domain (see Fig. \ref{Img_012}), these diagrams offer an immediate physical interpretation. We can appreciate, for example, that the width of the posterior predictive distributions varies as a function of the travel-time.}
\label{Img_013}
 \end{figure}

\begin{figure} 
  \centering
   \includegraphics[width=1\textwidth]{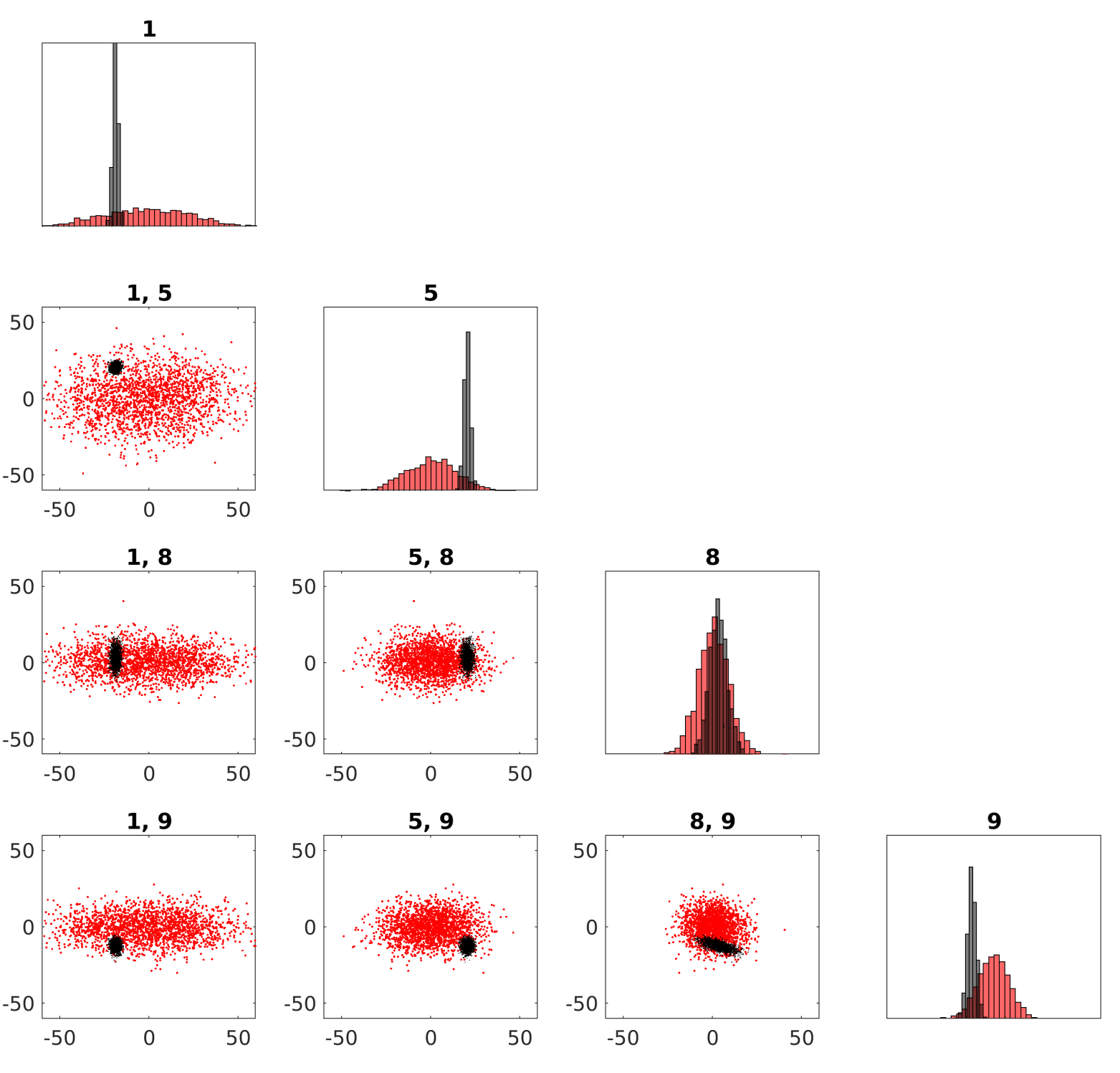}
\caption{Scatter plot representation of prior (red) and posterior (black) distributions for a set of input variables.}
\label{Img_014}
 \end{figure}

A more quantitative assessment can be achieved by comparing the reference input and the corresponding inversion solutions in terms of RMSE, Structural SIMilarity and Scoring values in the input domain \citep{gneiting2007strictly,levy2021using}.
We consider the posterior mean from the posterior distribution, including elements from the null space, and compute the RMSE and Structural SSIMilarity for inversion based on PCE and straight-line-approximation trained on $900$ sets with and without modelling error. To further assess both the statistical consistency between predictions and observations (calibration) and the sharpness of the prediction, we use the MCMC samples, transformed back in the pixel domain, to obtain a Gaussian approximation $\hat{P}$ of the estimated posterior PDF and use it to compute the logarithmic score with respect to the true value $\vu_{true}$ at each point, that is, $logS(\hat{P},\vu_{true})=-log(\hat{P}(\vu_{true}))$. Low values of the logarithmic score are associated with PDFs under which the true value has high probability \citep{good1992rational,friedli2021lithological}. Scoring values for inversion results associated with PCE and straight-line-approximation solvers including and ignoring model error are shown in Fig. \ref{SCORING}. 
The results of the fitting performances of the inversion results in the input and output domains are summarized in Table \ref{table:1}.
\begin{table}
\begin{center}
\caption{Assessment of the inversion results in the input and output domains for PCE and straight-line-approximation-based inversion, with or without modelling error included in the definition of the likelihood function. }
\begin{tabular}{||c c c c c c||} 
 \hline
 Model & RMSE Mean $\epsilon_r$  & SSIM Mean $\epsilon_r$ & Mean LogS $\epsilon_r$ & rmse  Mean PCE Output & rmse  Mean Gather Output \\ [0.5ex] 
 \hline\hline
 PCE & 1.10 & 0.69 & 1.63 & 1.07 ns &  1.17 ns\\ 
 \hline
 PCE NO & 1.09 & 0.69 & 1.63 & 1.10 ns & 1.19 ns\\
 \hline
 SL & 1.23 & 0.66 & 2.13 & 1.54 ns & 1.26 ns \\
 \hline
 SL NO & 1.56 & 0.63 & 3.72 & 4.70 ns & 2.05 ns \\ [1ex] 
 \hline
\end{tabular}
\label{table:1}
\end{center}
\end{table}
PCE-based inversions, with or without modelling error (PCE and PCE NO in Table \ref{table:1}, respectively) undeniably perform better than the corresponding schemes using the straight-line-approximation (SL and SL NO, respectively). The difference is substantial when RMSE in the input domain is considered, while the discrepancy becomes less important when Structural SIMilarity is used as a measure. Note that, as also observed by \citet{levy2021using}, including the modelling error in the inversion has as expected a minor adverse effect on fitting in the input domain regardless of the modelling scheme used as a consequence of the more inflated likelihood function. Performances differ very clearly in favor of PCE when scoring values are considered. Both PCE results (Figs. \ref{SCORING}(a-b)) provide significantly smaller values of the logarithmic score in large portions of the domain with respect to the straight-line solutions (Figs. \ref{SCORING}(c-d)). The differences in-between the PCE results are very minor, indicating only an extremely small improvement when modelling error is taken into account (see yellow arrows in Figs. \ref{SCORING}(b)). By contrast, including modelling errors has a dramatic and beneficial impact on the straight-line solutions (see white arrows in Fig. \ref{SCORING}(d) and compare with Fig. \ref{SCORING}(c)).
From here onward, we exclusively consider the PCE-results associated with $900$ training sets including modelling error (see Fig. \ref{Img_009b}(c)) as our reference inversion result. For these results, we show posterior realizations based on the reduced model parameterizations (Figs. \ref{Img_011}a-e). These images are very smoothly varying property fields and they do not express the posterior variability of the original problem. In contrast, Figs. \ref{Img_011}f-j show samples from the full posterior distribution obtained by including elements from the assumed null space (that is from the prior distribution of the higher-order truncated PCA coefficients). We see that these latter realizations bring in small-scale variability that can be highly important in subsurface applications particularly with respect to flow- and transport applications, but that are unresolvable by the data. 
The quality of the inversion results can be further analysed by looking at PCE evaluations on ensembles consisting of a number of random realizations from the prior and the posterior distributions. The resulting distributions obtained by applying the PCE on the prior and posterior samples (known as prior and posterior predictive distributions) allows us to appreciate the reduction in uncertainty in the output domain of the posterior with respect to the prior.
In this case we consider $10000$  realizations both of the prior and posterior, and summarize the corresponding output values in Fig. \ref{Img_012} along the observed data and the PCE evaluated at the posterior mean.
Not only is the match between the predictions based on the posterior mean and the observed data good, but we can also appreciate the much thinner posterior predictive distributions compared to the prior predictive distributions. %Especially for some of the output variables in Fig. \ref{Img_010} the difference between the posterior and predictive data are significant.
This analysis in the PCA output domain does not offer an immediate physical interpretation, but we can transform these predictions back into the more natural Gather output domain. This is done in Fig. \ref{Img_013} showing, for each source/receiver pair considered the much smaller width of the posterior compared with the prior predictive distributions. Finally, we explore the inversion results in the PCA-based input domain be considering scatter plots for a number of representative input principal components (see Fig. \ref{Img_014}). 
The data set is again shown to be highly informative in reducing posterior with respect to prior uncertainty. Nevertheless, we can also see that some of the inverted coefficients, for example, the $8^{th}$ (see Fig. \ref{Img_002}(h)), are not well resolved. This is not a consequence of the PCE, but of the intrinsic limitations of borehole configurations, allowing for compensation between high and low slowness values in explaining travel-times and, therefore, resulting in low sensitivity with respect to certain principal components. Compared to a deterministic inversion procedure, these limitations are expressed in terms of quantifiable uncertainty and can be interrogated as such. 
\section{Discussion}

A new parametrization of the inverse problem can be attractive, even if forward solvers are applied on classical grids. One previous example of such an approach for GPR inversion in terms of hydrogeological applications is  \citet{giannakis2021fractal}. In their work,  they use randomly generated anisotropic 2D fractals to populate a training set, whose principal components are used to approximate fractal and non-fractal soil distributions. A heuristic FDTD algorithm, subsequently minimizing a full-waveform cost function along a few tens of ordered principal components, was shown to provide reliable and relatively inexpensive results by operating via FDTD on a massively reduced inversion domain.
In our case, despite the strong decrease in complexity offered by a PCA-based parametrization, using the FDTD solver on the reduced input domain would involve extremely demanding computational costs and MCMC inversion would still be unpractical. 
We have circumvented this limitation by coupling an appropriate prior-informed parameterisation with surrogate modelling.\\
Surrogate modelling has emerged as a computational paradigm over the last $20$ years. The massive decrease of the computational effort associated with such modelling is achieved by approximating the underlying computational model with a simple and easy-to-evaluate function. 
A key requirement to allow implementation of PCE is that the number of input variables describing the computational model is relatively small (i.e., up to few tens). As the number of input variables increases, the number of coefficients required to grows polynomially in both the size of the input and the maximum degree of the expansion, leading to non tractable problems if the model response is highly nonlinear in its input parameters \citep{schobi2014combining}.
In this contribution, we have used PCA to explore the effective dimensionality of the input, and this allowed us to successfully employ PCE. While a re-parametrization of the input dimension is essential, this could be achieved by other means than PCA. Other parametrizations (wavelets, Fourier coefficients, Slepian functions, spherical harmonics), tailored for specific applications, may be employed to express the effective dimensionality of the input domain. One setting in which PCE could be particularly effective, is in studies where only few key parameters (such as position, shape and basic physical properties of scatterers or layers) properly characterize the target medium. Once a parametrization has been found to reduce the effective dimensionality of the input domain and a high fidelity PCE has been designed, inexpensive and effective MCMC inversion can be implemented.
We have also shown how to include the impact of the ignored variables approximating the 'null space' in the likelihood function and in expressing the full posterior uncertainty.
We have relied on a standard Metropolis-Hastings sampling algorithm, but alternative more effective sampling methods, such as Adaptive Metropolis, Hamiltonian Monte Carlo, or the Affine Invariant Ensemble Algorithm \citep{duane1987hybrid,haario2001adaptive,goodman2010ensemble} could be easily implemented in our workflow.\\
When considering modelling error in the likelihood function, our results always provided converging solutions with respect to Gelman-Rubin diagnostics, both for PCE- and straight-line-based inversion. Neglecting modeling error had a highly negative impact on inversions based on the (biased) straight-line approach in terms of RMSE, SSIM and LogS. The results obtained when including modeling errors in the likelihood function still provided less good results than the PCE-based forward modeling, most likely due to a larger variance and a skewed distributions implying that the Gaussian assumption in the likelihood is questionable. The effect of including the model error in the PCE-based inversion was very small, probably due to its unbiased nature, the relatively small variance and the Gaussian-like and symmetric error statistics. 
Most of the computational costs associated with MCMC inversion based on PCE is due to the training of the PCE surrogate, which in our case required the computation of several hundreds of FDTD data sets. However, the re-training of a PCE model is only needed when a different combination of forward model, prior model or experimental set-up are required. Also, it should be noted that these training sets of up to 1000 data sets is very small compared to what is used in deep-learning-based proxy modeling.
This makes this surrogate-modeling-based approach particularly attractive, for example, for monitoring applications where both solver and prior information remain unchanged between experimental campaigns.
Even if this was not necessary for the tests presented herein, re-training of a PCE may be needed during an MCMC inversion when considering more non-linear problems.
In travel-time computation, the degree of non-linearity of the forward problem depends largely on the variance of the slowness field. For the class of permittivity distributions we considered here, the straight-line-approximation is poor (see Fig. \ref{Img_009b}(d)) indicating a certain degree of non-linearity. Even so, a single surrogate model provided good results on input samples realistically representing the entire class of permittivity distributions involved (see Fig. \ref{Img_007}c). 
Therefore, a single PCE was used during the entire inversion process. However, when the forward model is more non-linear, constructing a globally-valid surrogate applicable for any realization of the prior can be challenging. In these cases, the accuracy of the surrogate can be refined in regions of high posterior probability \citep{li2014adaptive,wagner2021bayesian}.
Even if we considered borehole GPR applications herein, the the strategy could be easily adapted to other imaging problems such as active or passive seismic tomography at different scales \citep{bodin2009seismic,galetti2017transdimensional}.\\
Finally, a note on computational cost. All the numerical experiments discussed in this paper have been executed on a workstation equipped with 16GB DDR4 of RAM and powered by a 3.5GHz Quad-Core processor running matlab on Linux. With this configuration, the 15 chains / $1.5\times10^5$ iterations MCMC simulations presented here took around 6 hours to be executed. These performances indicate that PCE has the potential to allow MCMC inversion to be carried out without relying on massive computational resources.

\section{Conclusions}

%Parameterizations defined in terms of the grids used by standard forward solvers tend to neglect the global information content provided by prior knowledge. In standard FDTD methods, for example, the grid-size is determined by local maximum and minimum velocity values regardless of any other feature of the velocity distribution. Local properties and the corresponding parameters, however, may have limited effects in the output (data) domain. 
Model parameterizations for MCMC-based inversions are ideally informed by prior knowledge, as this allows for massive dimensionality reductions compared with parameterization based on the grids used in standard forward simulations. In a crosshole GPR setting, we have demonstrated how travel-time modelling using 30-50 prior-knowledge-based parameters based on PCA decomposition, instead of more than 30'000 parameters if relying on the FDTD grid, can predict the output well below the expected observational noise-level. We also demonstrate that high-quality PCA-based decomposition can be achieved in the output domain (from 289 travel-times to some tenths of PCA-coefficients). This massive reduction of the effective size of the input and output domain allows for efficient implementation of a PCE surrogate model using less than 1000 training sets. The resulting PCE surrogate model outperforms a straight-line-based solver that produces biased, higher variance and skewed model predictions. More importantly, the associated PCE-based modeling errors with respect to a FDTD solver are much smaller than the typical observational errors. Thus, allowing efficient replacement of FDTD solvers in the context of crosshole GPR travel-time tomography. This model reduction in the input and output domains combined with PCE-based modeling have dramatically reduced the computational burden of our problem, allowing us to successfully perform non-linear MCMC inversion at a low computational cost. Even if the PCE-based modeling error is low, we show how to account for it in the likelihood function and demonstrate that this inclusion is needed to obtain converging MCMC chains. We further highlight that the truncated higher-order PCA components that are ignored in the the dimension-reduced inversion need to be re-introduced when assessing posterior uncertainty and for obtaining realistic-looking posterior realizations. Indeed, even if the small-scale features cannot be resolved by the data, thereby, allowing them to be left out of the inversion process, they must be reintroduced to solve the original inverse problem (not letting the posterior distribution be affected by the dimensionality reduction) and also as small-scale unresolvable features might be important when using the results for predictions in terms of, for instance, hydrogeophysical applications involving transport processes that are highly impacted by small-scale heterogeneity. We stress that these results are expected to be easily extendable to other types of wave-based physics (e.g., seismics) also at larger-scales or in 3D, while applications to diffusion-based geophysics (e.g., electrical resistance tomography) would need to be investigated in detail in future studies.

\section*{Appendix A}

Compression in the output domain is convenient when considering PCE-based surogate modeling, as every output requires its own expansion which, in the MCMC sampling, is computed independently from the others. A reduction in output dimension imply, therefore, a decrease in computational costs and memory requirements, which can be particularly important when very long MCMC chains are used.
However, switching to a different coordinate set poses the problem of determining the statistical nature of the new output variables. We demonstrate that the complete and reduced output variables share the same relevant properties; this is a consequence of the linearity of PCA.
In the following, $N$ and $n$ represent the dimension of the complete and reduced output domains (in our considered example, $N$=289, $n$=35).
The standard output domain of travel-times, indicated here as $W_i, i \in [1, N]$, are assumed to constitute a set of independent identically-distributed Gaussian variables characterized by $\sigma^2=1 ns^2$.
The new set $Y_j, j \in [1, n]$ of output variables consists by definition of linear combinations of the random variables $W_i$, that is:
\begin{equation}
    Y_j = a_{ij} (W_i - \bar{W}_i), j \in [1, n],
\end{equation}
 where $\bar{W}_i$ and $a_{ij}$ stand for the sample mean and the elements of the orthogonal matrix representing the principal components, respectively (see Fig. \ref{Img_004}), and summation over repeated indices is implied. Each $Y_j, j \in [1, n]$ is given by the sum of Gaussian random variables and is, therefore, a Gaussian random variable.
We now consider the covariance operator associated with $Y_j, j \in [1, n]$:
\begin{equation}
    \mathrm{Cov}(Y_j,Y_k) = \mathrm{Cov}(a_{ij} (W_i - \bar{W}_i),a_{pk}(W_p - \bar{W}_p)).
\end{equation}
Invoking the bilinearity of the covariance operator, this reduces to:
\begin{equation}
\begin{split}
    \mathrm{Cov}(Y_j,Y_k) &= a_{ij}a_{pk} \mathrm{Cov}(W_i - \bar{W}_i,W_p - \bar{W}_p) \\
     &= a_{ij}a_{pk} \mathrm{Cov}(W_i,W_p) \\
     &= a_{ij}a_{pk} \sigma^2 \delta_i^p \\
     &= \delta_j^k \sigma^2  \\
    \end{split}
    \end{equation}
where $\delta_j^k$  is the Kronecker delta, which is equal to $1$ when $j=k$ and $0$ otherwise, and the last passage is justified by the orthogonality of the matrix $a_{ij}$, which is characterized by orthonormal columns/rows.
The new variables then constitute a set of i.i.d. Gaussian variables and can be treated as such in the Bayesian inversion process, notably in the definition of the likelihood function.

\clearpage

%%%%%%%%%%%%%%%%%%%%%%%%%%%%%%%%%%%%%%%%%%%%%%%%%%%%%%%%%%%%%%%%%%%%%%%%%%%%%%%%%%%%%%%%%%%%%%%%%%%%%%%
% ADD BIBLIOGRAPHY
%%%%%%%%%%%%%%%%%%%%%%%%%%%%%%%%%%%%%%%%%%%%%%%%%%%%%%%%%%%%%%%%%%%%%%%%%%%%%%%%%%%%%%%%%%%%%%%%%%%%%%%
\clearpage
\def\newblock{\ }
\bibliographystyle{apalike}
\bibliography{paper}

\providecommand{\noopsort}[1]{}\providecommand{\singleletter}[1]{#1}%
\begin{thebibliography}{}

\bibitem[Abbasi and Gholami, 2017]{abbasi2017polynomial}
Abbasi, M. and Gholami, A. (2017).
\newblock Polynomial chaos expansion for nonlinear geophysical inverse
  problems.
\newblock {\em Geophysics}, 82(4):R259--R268.

\bibitem[Annan, 2005]{annan2005gpr}
Annan, A.~P. (2005).
\newblock {GPR} methods for hydrogeological studies.
\newblock In {\em Hydrogeophysics}, pages 185--213. Springer.

\bibitem[Arcone et~al., 1998]{arcone1998ground}
Arcone, S.~A., Lawson, D.~E., Delaney, A.~J., Strasser, J.~C., and Strasser,
  J.~D. (1998).
\newblock Ground-penetrating radar reflection profiling of groundwater and
  bedrock in an area of discontinuous permafrost.
\newblock {\em Geophysics}, 63(5):1573--1584.

\bibitem[Balanis, 2012]{balanis2012advanced}
Balanis, C.~A. (2012).
\newblock {\em Advanced {E}ngineering {E}lectromagnetics}.
\newblock John Wiley \& Sons.

\bibitem[Blatman and Sudret, 2011]{blatman2011adaptive}
Blatman, G. and Sudret, B. (2011).
\newblock Adaptive sparse polynomial chaos expansion based on least angle
  regression.
\newblock {\em Journal of computational Physics}, 230(6):2345--2367.

\bibitem[Bodin and Sambridge, 2009]{bodin2009seismic}
Bodin, T. and Sambridge, M. (2009).
\newblock Seismic tomography with the reversible jump algorithm.
\newblock {\em Geophysical Journal International}, 178(3):1411--1436.

\bibitem[Cordua et~al., 2012]{cordua2012monte}
Cordua, K.~S., Hansen, T.~M., and Mosegaard, K. (2012).
\newblock Monte {C}arlo full-waveform inversion of crosshole {GPR} data using
  multiple-point geostatistical a priori information.
\newblock {\em Geophysics}, 77(2):H19--H31.

\bibitem[Dietrich and Newsam, 1997]{dietrich1997fast}
Dietrich, C.~R. and Newsam, G.~N. (1997).
\newblock Fast and exact simulation of stationary gaussian processes through
  circulant embedding of the covariance matrix.
\newblock {\em SIAM Journal on Scientific Computing}, 18(4):1088--1107.

\bibitem[Dou et~al., 2016]{dou2016real}
Dou, Q., Wei, L., Magee, D.~R., and Cohn, A.~G. (2016).
\newblock Real-time hyperbola recognition and fitting in {GPR} data.
\newblock {\em IEEE Transactions on Geoscience and Remote Sensing},
  55(1):51--62.

\bibitem[Duane et~al., 1987]{duane1987hybrid}
Duane, S., Kennedy, A.~D., Pendleton, B.~J., and Roweth, D. (1987).
\newblock Hybrid {M}onte {C}arlo.
\newblock {\em Physics letters B}, 195(2):216--222.

\bibitem[Ernst et~al., 2007]{ernst2007full}
Ernst, J.~R., Maurer, H., Green, A.~G., and Holliger, K. (2007).
\newblock Full-waveform inversion of crosshole radar data based on 2-{D}
  finite-difference time-domain solutions of {M}axwell's equations.
\newblock {\em IEEE transactions on geoscience and remote sensing},
  45(9):2807--2828.

\bibitem[Friedli et~al., 2021]{friedli2021lithological}
Friedli, L., Linde, N., Ginsbourger, D., and Doucet, A. (2021).
\newblock Lithological tomography with the correlated pseudo-marginal method.
\newblock {\em Geophysical Journal International}, 228(2):839--856.

\bibitem[Galetti et~al., 2017]{galetti2017transdimensional}
Galetti, E., Curtis, A., Baptie, B., Jenkins, D., and Nicolson, H. (2017).
\newblock Transdimensional {L}ove-wave tomography of the {B}ritish {I}sles and
  shear-velocity structure of the {E}ast {I}rish {S}ea {B}asin from
  ambient-noise interferometry.
\newblock {\em Geophysical Journal International}, 208(1):36--58.

\bibitem[Gelman and Rubin, 1992]{gelman1992inference}
Gelman, A. and Rubin, D.~B. (1992).
\newblock Inference from iterative simulation using multiple sequences.
\newblock {\em Statistical science}, 7(4):457--472.

\bibitem[Giannakis et~al., 2019]{giannakis2019machine}
Giannakis, I., Giannopoulos, A., and Warren, C. (2019).
\newblock A machine learning-based fast-forward solver for ground penetrating
  radar with application to full-waveform inversion.
\newblock {\em IEEE Transactions on Geoscience and Remote Sensing},
  57(7):4417--4426.

\bibitem[Giannakis et~al., 2021]{giannakis2021fractal}
Giannakis, I., Giannopoulos, A., Warren, C., and Sofroniou, A. (2021).
\newblock Fractal-constrained crosshole/borehole-to-surface full-waveform
  inversion for hydrogeological applications using ground-penetrating radar.
\newblock {\em IEEE Transactions on Geoscience and Remote Sensing}.

\bibitem[Gneiting and Raftery, 2007]{gneiting2007strictly}
Gneiting, T. and Raftery, A.~E. (2007).
\newblock Strictly proper scoring rules, prediction, and estimation.
\newblock {\em Journal of the American statistical Association},
  102(477):359--378.

\bibitem[Good, 1992]{good1992rational}
Good, I.~J. (1992).
\newblock Rational decisions.
\newblock In {\em Breakthroughs in statistics}, pages 365--377. Springer.

\bibitem[Goodman and Weare, 2010]{goodman2010ensemble}
Goodman, J. and Weare, J. (2010).
\newblock Ensemble samplers with affine invariance.
\newblock {\em Communications in applied mathematics and computational
  science}, 5(1):65--80.

\bibitem[Grasmueck et~al., 2005]{grasmueck2005full}
Grasmueck, M., Weger, R., and Horstmeyer, H. (2005).
\newblock Full-resolution 3{D} {GPR} imaging.
\newblock {\em Geophysics}, 70(1):K12--K19.

\bibitem[Haario et~al., 2004]{haario2004markov}
Haario, H., Laine, M., Lehtinen, M., Saksman, E., and Tamminen, J. (2004).
\newblock Markov chain {M}onte {C}arlo methods for high dimensional inversion
  in remote sensing.
\newblock {\em Journal of the Royal Statistical Society: series B (statistical
  methodology)}, 66(3):591--607.

\bibitem[Haario et~al., 2001]{haario2001adaptive}
Haario, H., Saksman, E., and Tamminen, J. (2001).
\newblock An adaptive {M}etropolis algorithm.
\newblock {\em Bernoulli}, pages 223--242.

\bibitem[Hansen and Cordua, 2017]{hansen2017efficient}
Hansen, T.~M. and Cordua, K.~S. (2017).
\newblock Efficient monte carlo sampling of inverse problems using a neural
  network-based forward—applied to gpr crosshole traveltime inversion.
\newblock {\em Geophysical Journal International}, 211(3):1524--1533.

\bibitem[Hansen et~al., 2014]{hansen2014accounting}
Hansen, T.~M., Cordua, K.~S., Jacobsen, B.~H., and Mosegaard, K. (2014).
\newblock Accounting for imperfect forward modeling in geophysical inverse
  problems—exemplified for crosshole tomography.
\newblock {\em Geophysics}, 79(3):H1--H21.

\bibitem[Hastings, 1970]{hastings1970monte}
Hastings, W.~K. (1970).
\newblock Monte {C}arlo sampling methods using {M}arkov chains and their
  applications.

\bibitem[Higdon et~al., 2015]{higdon2015bayesian}
Higdon, D., McDonnell, J.~D., Schunck, N., Sarich, J., and Wild, S.~M. (2015).
\newblock A {B}ayesian approach for parameter estimation and prediction using a
  computationally intensive model.
\newblock {\em Journal of Physics G: Nuclear and Particle Physics},
  42(3):034009.

\bibitem[Hunziker et~al., 2019]{hunziker2019bayesian}
Hunziker, J., Laloy, E., and Linde, N. (2019).
\newblock Bayesian full-waveform tomography with application to crosshole
  ground penetrating radar data.
\newblock {\em Geophysical Journal International}, 218(2):913--931.

\bibitem[Irving and Knight, 2006]{irving2006numerical}
Irving, J. and Knight, R. (2006).
\newblock Numerical modeling of ground-penetrating radar in 2-{D} using
  {MATLAB}.
\newblock {\em Computers \& Geosciences}, 32(9):1247--1258.

\bibitem[Khan et~al., 2016]{khan2016single}
Khan, A., van Driel, M., B{\"o}se, M., Giardini, D., Ceylan, S., Yan, J.,
  Clinton, J., Euchner, F., Lognonn{\'e}, P., Murdoch, N., et~al. (2016).
\newblock Single-station and single-event marsquake location and inversion for
  structure using synthetic martian waveforms.
\newblock {\em Physics of the Earth and Planetary Interiors}, 258:28--42.

\bibitem[Kuroda et~al., 2007]{Kuroda2007}
Kuroda, S., Takeuchi, M., and Kim, H.~J. (2007).
\newblock {Full-waveform inversion algorithm for interpreting crosshole radar
  data: a theoretical approach}.
\newblock {\em Geosciences Journal}, 11(3):211--217.

\bibitem[LaBrecque et~al., 2002]{labrecque2002three}
LaBrecque, D., Alumbaugh, D.~L., Yang, X., Paprocki, L., and Brainard, J.
  (2002).
\newblock Three-dimensional monitoring of vadose zone infiltration using
  electrical resistivity tomography and cross-borehole ground-penetrating
  radar.
\newblock In {\em Methods in Geochemistry and Geophysics}, volume~35, pages
  259--272. Elsevier.

\bibitem[Laloy et~al., 2015]{laloy2015probabilistic}
Laloy, E., Linde, N., Jacques, D., and Vrugt, J.~A. (2015).
\newblock Probabilistic inference of multi-{G}aussian fields from indirect
  hydrological data using circulant embedding and dimensionality reduction.
\newblock {\em Water Resources Research}, 51(6):4224--4243.

\bibitem[Laloy et~al., 2013]{laloy2013efficient}
Laloy, E., Rogiers, B., Vrugt, J.~A., Mallants, D., and Jacques, D. (2013).
\newblock Efficient posterior exploration of a high-dimensional groundwater
  model from two-stage {M}arkov chain {M}onte {C}arlo simulation and polynomial
  chaos expansion.
\newblock {\em Water Resources Research}, 49(5):2664--2682.

\bibitem[Levy et~al., 2021]{levy2021using}
Levy, S., Hunziker, J., Laloy, E., Irving, J., and Linde, N. (2021).
\newblock Using deep generative neural networks to account for model errors in
  markov chain monte carlo inversion.
\newblock {\em Geophysical Journal International}, 228(2):1098--1118.

\bibitem[Li and Marzouk, 2014]{li2014adaptive}
Li, J. and Marzouk, Y.~M. (2014).
\newblock Adaptive construction of surrogates for the {B}ayesian solution of
  inverse problems.
\newblock {\em SIAM Journal on Scientific Computing}, 36(3):A1163--A1186.

\bibitem[L\"uthen et~al., 2021a]{luethen2021automatic}
L\"uthen, N., Marelli, S., and Sudret, B. (2021a).
\newblock Automatic selection of basis-adaptive sparse polynomial chaos
  expansions for engineering applications.

\bibitem[L\"uthen et~al., 2021b]{luethen2021sparse}
L\"uthen, N., Marelli, S., and Sudret, B. (2021b).
\newblock Sparse polynomial chaos expansions: Literature survey and benchmark.
\newblock {\em SIAM/ASA Journal on Uncertainty Quantification}, 9(2):593--649.

\bibitem[Madsen and Hansen, 2018]{madsen2018estimation}
Madsen, R.~B. and Hansen, T.~M. (2018).
\newblock Estimation and accounting for the modeling error in probabilistic
  linearized amplitude variation with offset inversion.
\newblock {\em Geophysics}, 83(2):N15--N30.

\bibitem[Marelli et~al., 2021a]{UQdoc_14_104}
Marelli, S., L\"uthen, N., and Sudret, B. (2021a).
\newblock {UQLab user manual -- Polynomial chaos expansions}.
\newblock Technical report, Chair of Risk, Safety and Uncertainty
  Quantification, ETH Zurich,Switzerland.
\newblock Report \# UQLab-V1.4-104.

\bibitem[Marelli and Sudret, 2014]{marelli2014uqlab}
Marelli, S. and Sudret, B. (2014).
\newblock {UQLab}: A framework for uncertainty quantification in {M}atlab.
\newblock In {\em Vulnerability, uncertainty, and risk: quantification,
  mitigation, and management}, pages 2554--2563.

\bibitem[Marelli et~al., 2021b]{marelli2021stochastic}
Marelli, S., Wagner, P.-R., Lataniotis, C., and Sudret, B. (2021b).
\newblock Stochastic spectral embedding.
\newblock {\em International Journal for Uncertainty Quantification}, 11(2).

\bibitem[Marzouk and Xiu, 2009]{marzouk2009stochastic}
Marzouk, Y. and Xiu, D. (2009).
\newblock A stochastic collocation approach to {B}ayesian inference in inverse
  problems.

\bibitem[Marzouk et~al., 2007]{marzouk2007stochastic}
Marzouk, Y.~M., Najm, H.~N., and Rahn, L.~A. (2007).
\newblock Stochastic spectral methods for efficient {B}ayesian solution of
  inverse problems.
\newblock {\em Journal of Computational Physics}, 224(2):560--586.

\bibitem[M{\'e}tivier et~al., 2020]{metivier2020efficient}
M{\'e}tivier, D., Vuffray, M., and Misra, S. (2020).
\newblock Efficient polynomial chaos expansion for uncertainty quantification
  in power systems.
\newblock {\em Electric Power Systems Research}, 189:106791.

\bibitem[Nagel, 2019]{nagel2019bayesian}
Nagel, J.~B. (2019).
\newblock Bayesian techniques for inverse uncertainty quantification.
\newblock {\em IBK Bericht}, 504.

\bibitem[Nielsen et~al., 2010]{nielsen2010estimation}
Nielsen, L., Looms, M.~C., Hansen, T.~M., Cordua, K.~S., Stemmerik, L., Miller,
  R., Bradford, J., and Holliger, K. (2010).
\newblock Estimation of chalk heterogeneity from stochastic modeling
  conditioned by crosshole {GPR} traveltimes and log data.
\newblock {\em Advances in near-surface seismology and ground-penetrating
  radar: SEG Geophysical Development Series}, 15:379--398.

\bibitem[Olsson et~al., 1992]{olsson1992borehole}
Olsson, O., Falk, L., Forslund, O., Lundmark, L., and Sandberg, E. (1992).
\newblock Borehole radar applied to the characterization of hydraulically
  conductive fracture zones in crystalline rock.
\newblock {\em Geophysical prospecting}, 40(2):109--142.

\bibitem[Piscitelli et~al., 2007]{piscitelli2007gpr}
Piscitelli, S., Rizzo, E., Cristallo, F., Lapenna, V., Crocco, L., Persico, R.,
  and Soldovieri, F. (2007).
\newblock {GPR} and microwave tomography for detecting shallow cavities in the
  historical area of “{S}assi of {M}atera”(southern {I}taly).
\newblock {\em Near Surface Geophysics}, 5(4):275--284.

\bibitem[Rasmussen, 2003]{rasmussen2003gaussian}
Rasmussen, C.~E. (2003).
\newblock Gaussian processes in machine learning.
\newblock In {\em Summer school on machine learning}, pages 63--71. Springer.

\bibitem[Rawlinson and Urvoy, 2006]{rawlinson2006simultaneous}
Rawlinson, N. and Urvoy, M. (2006).
\newblock Simultaneous inversion of active and passive source datasets for
  3-{D} seismic structure with application to {T}asmania.
\newblock {\em Geophysical Research Letters}, 33(24).

\bibitem[Sacks et~al., 1989]{sacks1989designs}
Sacks, J., Schiller, S.~B., and Welch, W.~J. (1989).
\newblock Designs for computer experiments.
\newblock {\em Technometrics}, 31(1):41--47.

\bibitem[Sch{\"o}bi et~al., 2014]{schobi2014combining}
Sch{\"o}bi, R., Kersaudy, P., Sudret, B., and Wiart, J. (2014).
\newblock {\em Combining polynomial chaos expansions and kriging}.
\newblock PhD thesis, ETH Zurich, Switzerland; Orange Labs research.

\bibitem[Slob et~al., 2010]{slob2010surface}
Slob, E., Sato, M., and Olhoeft, G. (2010).
\newblock Surface and borehole ground-penetrating-radar developments.
\newblock {\em Geophysics}, 75(5):75A103--75A120.

\bibitem[Sochala et~al., 2021]{sochala2021polynomial}
Sochala, P., Gesret, A., and Le~Maitre, O. (2021).
\newblock Polynomial surrogates for bayesian traveltime tomography.
\newblock {\em GEM-International Journal on Geomathematics}, 12(1):1--34.

\bibitem[Stotzka et~al., 2002]{stotzka2002medical}
Stotzka, R., Wuerfel, J., Mueller, T.~O., and Gemmeke, H. (2002).
\newblock Medical imaging by ultrasound computer tomography.
\newblock In {\em Medical Imaging 2002: Ultrasonic Imaging and Signal
  Processing}, volume 4687, pages 110--119. International Society for Optics
  and Photonics.

\bibitem[Taillandier et~al., 2009]{taillandier2009first}
Taillandier, C., Noble, M., Chauris, H., and Calandra, H. (2009).
\newblock First-arrival traveltime tomography based on the adjoint-state
  method.
\newblock {\em Geophysics}, 74(6):WCB1--WCB10.

\bibitem[Tant et~al., 2018]{tant2018transdimensional}
Tant, K. M.~M., Galetti, E., Mulholland, A., Curtis, A., and Gachagan, A.
  (2018).
\newblock A transdimensional {B}ayesian approach to ultrasonic travel-time
  tomography for non-destructive testing.
\newblock {\em Inverse Problems}, 34(9):095002.

\bibitem[Torre et~al., 2019]{torre2019data}
Torre, E., Marelli, S., Embrechts, P., and Sudret, B. (2019).
\newblock Data-driven polynomial chaos expansion for machine learning
  regression.
\newblock {\em Journal of Computational Physics}, 388:601--623.

\bibitem[Vu et~al., 2020]{vu2020magnetometric}
Vu, M., Jardani, A., Revil, A., and Jessop, M. (2020).
\newblock Magnetometric resistivity tomography using chaos polynomial
  expansion.
\newblock {\em Geophysical Journal International}, 221(3):1469--1483.

\bibitem[Wagner et~al., 2020]{wagner2020bayesian}
Wagner, P.-R., Fahrni, R., Klippel, M., Frangi, A., and Sudret, B. (2020).
\newblock Bayesian calibration and sensitivity analysis of heat transfer models
  for fire insulation panels.
\newblock {\em Engineering Structures}, 205:110063.

\bibitem[Wagner et~al., 2021a]{wagner2021bayesian}
Wagner, P.-R., Marelli, S., and Sudret, B. (2021a).
\newblock Bayesian model inversion using stochastic spectral embedding.
\newblock {\em Journal of Computational Physics}, 436:110141.

\bibitem[Wagner et~al., 2021b]{UQdoc_14_113}
Wagner, P.-R., Nagel, J., Marelli, S., and Sudret, B. (2021b).
\newblock {UQLab user manual -- Bayesian inversion for model calibration and
  validation}.
\newblock Technical report, Chair of Risk, Safety and Uncertainty
  Quantification, ETH Zurich,Switzerland.
\newblock Report UQLab-V1.4-113.

\bibitem[Wang et~al., 2004]{wang2004image}
Wang, Z., Bovik, A.~C., Sheikh, H.~R., and Simoncelli, E.~P. (2004).
\newblock Image quality assessment: from error visibility to structural
  similarity.
\newblock {\em IEEE transactions on image processing}, 13(4):600--612.

\bibitem[Warren et~al., 2019]{warren2019cuda}
Warren, C., Giannopoulos, A., Gray, A., Giannakis, I., Patterson, A., Wetter,
  L., and Hamrah, A. (2019).
\newblock A {CUDA}-based {GPU} engine for {gprMax}: Open source {FDTD}
  electromagnetic simulation software.
\newblock {\em Computer Physics Communications}, 237:208--218.

\bibitem[Xiu and Karniadakis, 2002]{xiu2002wiener}
Xiu, D. and Karniadakis, G.~E. (2002).
\newblock The {W}iener-{A}skey polynomial chaos for stochastic differential
  equations.
\newblock {\em SIAM journal on scientific computing}, 24(2):619--644.

\bibitem[Zhao et~al., 2008]{zhao2008seismic}
Zhao, D., Lei, J., and Liu, L. (2008).
\newblock Seismic tomography of the {M}oon.
\newblock {\em Chinese Science Bulletin}, 53(24):3897--3907.

\end{thebibliography}

%\bibliographystyle{unsrtnat}
%\bibliography{references}  %%% Uncomment this line and comment out the ``thebibliography'' section below to use the external .bib file (using bibtex) .

\end{document}